\documentclass{aa}

\usepackage{hyperref}
\usepackage{cleveref}
\usepackage{graphicx}
\usepackage{txfonts}
\usepackage{indentfirst}
\usepackage{amsmath}
\usepackage{hyperref} 
\usepackage{amsmath,bm}
\usepackage{booktabs} 
\usepackage{floatrow}
\usepackage{amsmath,bm}

\usepackage[english]{babel}

\usepackage{natbib}

\usepackage[toc,page]{appendix}

\usepackage[nottoc]{tocbibind}

\DeclareFloatFont{tiny}{\tiny}
\usepackage[utf8]{inputenc}

\begin{document} 
        
\title{Constraining the detectability of water ice in debris disks} 

\author{M. Kim\inst{1}
\and
S. Wolf\inst{1} 
\and 
A. Potapov\inst{2}
\and 
H. Mutschke\inst{3}
\and 
C. J\"ager\inst{2}}

\institute{Institut f\"ur Theoretische Physik und Astrophysik, Christian-Albrechts-Universit\"at zu Kiel, Leibnizstra\ss e 15, 24118 Kiel, Germany\\
\email{mkim@astrophysik.uni-kiel.de}
\and 
Laborastrophysikgruppe des Max-Planck-Instituts f\"ur Astronomie am Institut f\"ur Festk\"orperphysik, Friedrich-Schiller-Universit\"at Jena, Helmholtzweg 3, 07743 Jena, Germany
\and 
Astrophysikalisches Institut und Universit\"ats-Sternwarte, Friedrich-Schiller-Universit\"at Jena, Schillerg\"a\ss chen 2-3, 07745 Jena, Germany}

\date{}

\titlerunning  {Feasibility of detecting water ice in debris disks}
\authorrunning {M. Kim et al.} 


  \abstract
   {Water ice is important for the evolution and preservation of life. Identifying the distribution of water ice in debris disks is therefore of great interest in the field of astrobiology. Furthermore, icy dust grains are expected to play important roles throughout the entire planet formation process. However, currently available observations only allow deriving weak conclusions about the existence of water ice in debris disks.}
   {We investigate whether it is feasible to detect water ice in typical debris disk systems. We take the following ice destruction mechanisms into account: sublimation of ice, dust production through planetesimal collisions, and photosputtering by UV-bright central stars. We consider icy dust mixture particles with various shapes consisting of amorphous ice, crystalline ice, astrosilicate, and vacuum inclusions (i.e., porous ice grains).}
   {We calculated optical properties of inhomogeneous icy dust mixtures using effective medium theories, that is, Maxwell-Garnett rules. Subsequently, we generated synthetic debris disk observables, such as spectral energy distributions and spatially resolved thermal reemission and scattered light intensity and polarization maps with our code DMS.}
   {We find that the prominent $\sim$ 3 $\mu\rm{m}$ and 44 $\mu\rm{m}$ water ice features can be potentially detected in future observations of debris disks with the James Webb Space Telescope (JWST) and the Space Infrared telescope for Cosmology and Astrophysics (SPICA). We show that the sublimation of ice, collisions between planetesimals, and photosputtering caused by UV sources clearly affect the observational appearance of debris disk systems. In addition, highly porous ice (or ice-rich aggregates) tends to produce highly polarized radiation at around 3 $\mu\rm{m}$. Finally, the location of the ice survival line is determined by various dust properties such as a fractional ratio of ice versus dust, physical states of ice (amorphous or crystalline), and the porosity of icy grains.}
   {}


   \maketitle


\section{Introduction}

\noindent Water ice (hereafter referred to as "ice") is assumed to play an important role during planet formation (\citealp{Thommes}; \citealp{Min}). The ice is thought to immediately sublimate in the hot inner regions of circumstellar disks, therefore it is expected to be present only beyond the ice sublimation front, the so-called snow line. Consequently, the formation of the planetary core in the core formation or gas capture scenario is significantly affected by the freeze-out of water onto dust grains (\citealp{Stevenson}; \citealp{Hubickyj}). Furthermore, at later stages of the formation and early evolution of planetary systems, icy planetesimals, icy pebbles, or cometary objects may deliver water to rocky planets (\citealp{Morbidelli}; \citealp{Raymond}; \citealp{Nagasawa}; \citealp{Woitkea, Woitkeb}) and also to the innermost part of the remaining disk (\citealp{Eisner}). Understanding the origin and transport of water to Earth finally is of key importance for deciphering the conditions during the early evolution of life.\newline    
\indent The OH stretching vibrational modes are active for cation-bonded hydroxyl groups within H$_{\rm 2}$O molecules (and minerals at the surface of refractory grains) around 3~$\mu\rm{m}$, which is active when H$_{\rm 2}$O is present as ice (\citealp{Beck}; \citealp{Whittet}). The observation of the 3~$\mu\rm{m}$ ice feature therefore is the main target for ice detection in astrophysical environments. In addition, transverse optical and longitudinal acoustic vibrational modes are active around 44~$\mu\rm{m}$ and 62~$\mu\rm{m}$, respectively (\citealp{Bertie}; \citealp{Omont}; \citealp{Smith1994}; \citealp{Dartois}). An observation of these ice features, for instance, 44~$\mu\rm{m}$ and 62~$\mu\rm{m}$, would therefore provide crucial evidence and possible constraints for the presence and properties of ice as well. \newline
\indent In various protoplanetary disks, the ice features at 3~$\mu\rm{m}$, 44~$\mu\rm{m}$, and 62~$\mu\rm{m}$ have been detected and analyzed (\citealp{Pontoppidan}; \citealp{Terada}; \citealp{Honda09};  \citealp{Schegerer}; \citealp{Aikawa}; \citealp{Molinari}; \citealp{Malfait}; \citealp{McClure12, McClure15}). However, only the detection of the 62~$\mu\rm{m}$ ice feature has so far been inferred in debris disks (\citealp{Chen}). Our observational understanding of the spatial distribution of icy grains is therefore in its infancy, and even the presence of ice in debris disks is hardly established observationally. \newline
\indent Next-generation~observatories~are~expected~to~allow~making significant progress on our understanding of the ice distribution in debris disks. Combined observations of JWST/NIRCam (James Webb Space Telescope/Near Infrared Camera; aiming for observations at wavelengths from 0.6~$\mu\rm{m}$~to~5~$\mu\rm{m}$, \citealp{STScI}) and ELT/METIS (Extremely Large Telescope/The Mid-Infrared E-ELT Imager and Spectrograph; aiming for~observations at wavelengths from 3~$\mu\rm{m}$ to 20~$\mu\rm{m}$, \citealp{Brandl}) are expected to play a leading role in confirming the presence (or absence) of ice in debris disks. In addition, SPICA/SAFARI (Space Infrared telescope for Cosmology and Astrophysics/SpicA FAR-infrared Instrument; aiming for observations at wavelengths from 34~$\mu\rm{m}$ to 230~$\mu\rm{m}$, \citealp{Jellema}) will potentially contribute to the understanding of ice in debris disks. \newline
\indent The goal of this study is to answer the key question of the observational requirements either to constrain the detectability of ice in debris disks or to provide useful limits for the existence, properties, and spatial distribution of ice in debris disks. For this purpose, we have conducted a numerical feasibility study assuming various fractional ratios of ice, porosities, and shapes of aggregates of icy dust mixtures in debris disks. Subsequently, we investigated whether selected instruments or observatories that will become available in the near future, such as the JWST/NIRCam and SPICA/SAFARI, will indeed allow contributing to answering this question. \newline
\indent This paper is organized as follows: In Section 2 we describe the underlying physics related to ice depletion in debris disks. In Section 3 we depict our typical reference debris disks model. In Section 4 we investigate the influence of icy dust parameters on the resulting spectral energy distributions (SEDs) and spatially resolved images of debris disks system. In addition, we predict the boundary of the ice reservoir that is referred to as the "ice survival line" in the following (in contrast to the term "snow line", which is used to characterize the region where gas begins to freeze out onto dust grains in protoplanetary gas-rich disks). We finally evaluate and constrain the detectability of icy dust grains with future observatories. We summarize our findings in Section 5. \newline


\section{Depletion of ice in debris disks}

{\hspace{-2mm}}\noindent \textbf{Sublimation}{\hspace{2mm}} If icy grains (at least partially) drift radially inward due to the Poynting-Robertson effect (\citealp{Poynting}; \citealp{Robertson}; hereafter referred to as P-R effect), they pile up and form a ring: their inward drift is suppressed by stellar radiation pressure when the ratio of radiation pressure to stellar gravity, that is, $\beta$ $\equiv$ $F_{\rm rp}/F_{\rm grav}$, on them increases during their sublimation phases as a result of decreasing particle mass loss (\citealp{Kobayashi2010}). Eventually, ice immediately sublimates when its temperatures reach the sublimation temperature. Ice sublimation is therefore considered as a possible explanation for the presence of central clearing in debris disks (\citealp{Jura}). The sublimation temperature depends on the gas pressure, which itself is a function of the evolutionary state of circumstellar disks (\citealp{Fraser}; \citealp{Collings}; \citealp{BrownBolina}; \citealp{Feistel}). In addition, mixing of ice with dust grains can alter the kinetics of ice desorption (\citealp{potapov20182}). The sublimation temperature is independent of dust sizes, therefore small hotter grains sublimate before cooler large grains do (\citealp{Kobayashi2008}). \newline
\indent  \citet{Jura} reproduced the IR emission detected in the HR 4796 system by~110 K blackbody grains. In addition, Spitzer IRS spectra around A, B, and F stars with IR excesses were analyzed and fit with a single-temperature~(110~-~120~K)~blackbody by \citet{Chen}. This absence of warmer grains could be interpreted as the result of the ice sublimation in the inner region.  \citet{Golimowski} interpreted the observed color change beyond 120~au in the archetypal $\beta$ Pic disks as a possible indication of ice sublimation, which may result in larger average grain sizes, that is, cooler grains, beyond the sublimation zone. \citet{Kobayashi2010} indicated that the flat radial profile of the dust flux at 10~-~50 au and at 5~-~15 au derived from in situ dust impacts measured with the Voyager and Pioneer spacecrafts, respectively, may be caused by ice sublimation. \newline

{\hspace{-2mm}}\noindent \textbf{UV photosputtering}{\hspace{2mm}} Individual UV photons absorbed by an ice grain do not only dissociate water molecules, but can cause OH to be directly desorbed from the surface of ice grain. Alternatively, the molecule recombines. This process is known as UV photosputtering (\citealp{Artymowicz}; \citealp{Dominik}; \citealp{Grigorieva}). Because debris disks are transparent to the stellar radiation, energetic UV photon can efficiently penetrate the disks out to very large distances. \citet{Brown} indicated that the UV photosputtering rate becomes higher than the sublimation rate beyond 5 AU in the solar system. In addition, \citet{Oka} found that far-UV photosputtering radiation depresses the ice-condensation front toward the mid-plane and pushes the surface ice snow line significantly outward. These studies imply that ice can be destroyed outside the sublimation distance as well.\newline 
\indent \citet{Grigorieva} predicted that UV photosputtering efficiently destroys ice in optically thin disks, even far beyond the ice sublimation line. This means that UV photosputtering is responsible for the internal structure, thereby further increasing the effective grain size. \citet{Loehne2012} estimated UV photosputtering lifetimes compared to collisional lifetimes of objects, so that this rough comparison shows that UV photosputtering cannot be a negligible removal mechanism for ice grains with radii smaller than a few tens of~$\mu\rm{m}$. Furthermore, the analysis of \textit{Herschel} observations shows that the resolved cold debris disks around HD 61005, HD 104860, and HD 107146 require a minimum grain $a$$_{\rm min}$ about five times larger than the blow-out size grains $a$$_{\rm BO}$ (\citealp{Morales}). This observation is indicative for an increase of the effective grain size in debris disks system through grain depletion by UV photosputtering. On the other hand, \citet{Honda16} detected a shallow 3~$\mu\rm{m}$ ice feature that might be caused by the UV photosputtering in Herbig Be HD100546 disk scattered light spectra. This would mean that UV photosputtering is responsible for the strength of the ice feature as well. \newline 
\indent \citet{Johnson} predicted that porosity of interstellar and circumstellar grains can significantly lower the photosputtering yield. In addition, recent laboratory experiments have demonstrated trapping of water molecules on porous silicate grains at 200\,K (above the desorption temperature of H$_{\rm 2}$O ice; \citealp{potapov20182}; \citealp{Potapov}). First experiments on the UV photosputtering of water ice molecules from the surface of porous silicate and carbon grains by UV photons showed an influence of the surface properties on the photosputtering yield, in particular in the monolayer regime (\citealp{potapov2019}). \newline

\begin{table*}[h!]
\def\arraystretch{1.2}                                                                  
\caption{Model parameters for the simulation of our reference debris disk model.}           
\label{table:1}                                                                         
\centering                                                                              
\begin{tabular}{l l}                                                                    
\hline\hline                                                                            
  Parameter                                                                             & Value \\
\hline
  Stellar type                                                                          & A6 V (\citealp{Gray})\\
  Mass of the star $M_{\rm *}$                                                          & 1.75\,$ \rm M_{\rm \odot}$ (\citealp{Kervella})\\
  Radius of the star $R_{\rm *}$                                                        & 1.8\,$ \rm R_{\rm \odot}$ (\citealp{Crifo})\\
  Effective temperature $T_{\rm *}$                                                     & 8052\,\rm K (\citealp{Gray}) \\
  Distance to the debris disk system $d$                                                & 19.3\,pc\\
  Inner radius of the debris disk system $R_{\rm in}$                                   & 3 \,au \\
  Outer radius of the debris disk system $R_{\rm out}$                                  & 150 \,au \\
  Radial density distribution $n(a)$                                                             & $n(r)$ $\propto$ $r^{\rm  - 1.5}$ (\citealp{Krivov2006}; \citealp{Strubbe})\\

  Disk inclination                                                                      & 0 $^{\circ}$ (face-on disk) \\
  Size range modeling $n(r)$                                                            & [0.1 $\rm {\mu m}$, 1000 $\rm {\mu m}$]  with $n(a)$ $\propto$ $a^{\rm  - 3.5}$ (\citealp{Dohnanyi})\\
  Dust composition and                                                                                   & Amorphous ice (\citealp{Potapov}, \citealp{Curtis}, and \citealp{Li})\normalsize\\
  References of corresponding optical data                                              & Crystalline ice (\citealp{Reinert}, \citealp{Haessner}, \citealp{Mishima},\\ 
                                                                                        & \citealp{Potapov}, \citealp{Warren}, \citealp{Curtis}, and \citealp{Li}) \normalsize\\
                                                                                        & Astrosil (\citealp{Draine})\normalsize\\
  Fractional ratio of ice ${\mathcal{F}}_{\rm ice}$ in icy dust mixtures                         & 0 (pure Astrosil), 0.25, 0.5, 0.75, and 1 (pure ice)\\
  Porosity of grains $\mathcal{P}$                                                               & 0 (compact), 0.25, 0.5, 0.75\\
  Sublimation temperature                                                               & 100\,K: Pure amorphous ice\\ 
  (\citealp{BrownR}; \citealp{Kobayashi2010})                                                                                   & 105\,K: Pure crystalline ice\\
                                                                                                                & 100\,K: Amorphous ice in dust aggregates\\
                                                                                                                & 105\,K: Crystalline ice in dust aggregates\\

\hline                                                                                  
\end{tabular}
\end{table*}

\begin{figure*}
\centering
\includegraphics[width=18cm]{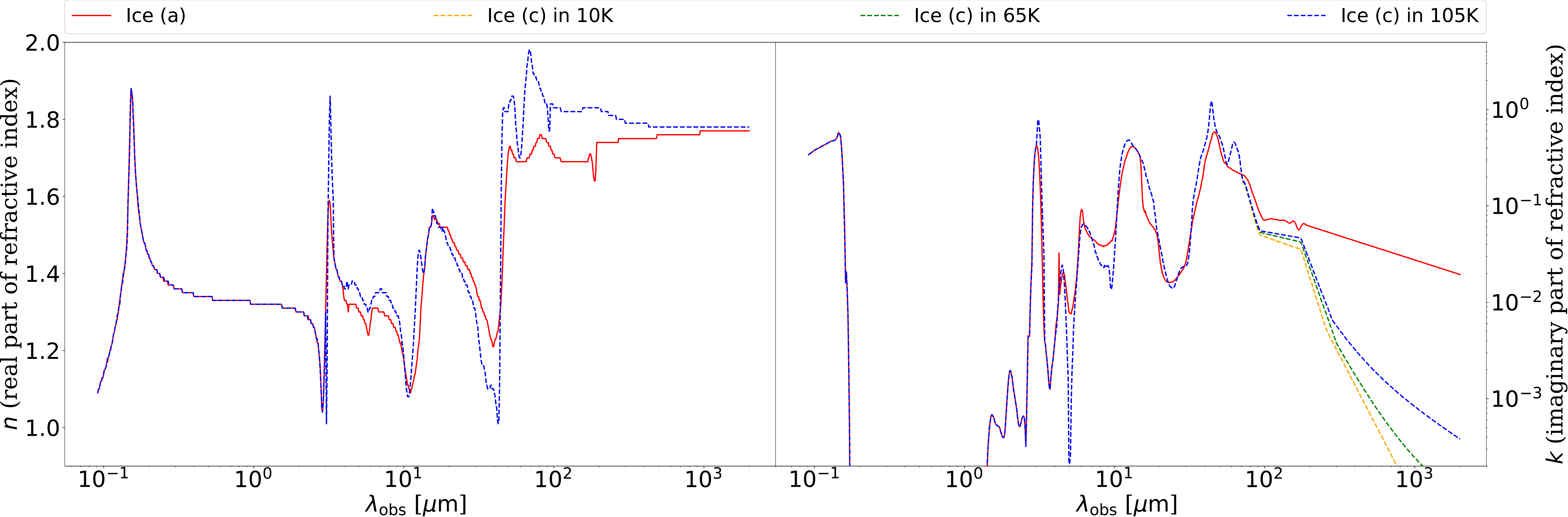}
\caption{Optical constants $n$ and $k$, i.e., the real and imaginary part of the refractive index, of pure amorphous ice and crystalline ice depending on the temperature. Ice (a) and ice (c) indicate amorphous ice (solid line) and crystalline ice (dashed line), respectively. Optical constants $n$ of ice (c) and $k$ of ice (c) in 0.1~$\mu\rm{m}$ to 62~$\mu\rm{m}$ show the same regardless of the temperature. However, $k$ of ice (c) at $\sim$ 62~$\mu\rm{m}$ to 1000~$\mu\rm{m}$ shows differences that sensitively depend on the temperature. A more detailed description can be found in Sect. 3.}
\label{FigVibStab}
\end{figure*}
\begin{figure*}
\centering
\includegraphics[width=18cm]{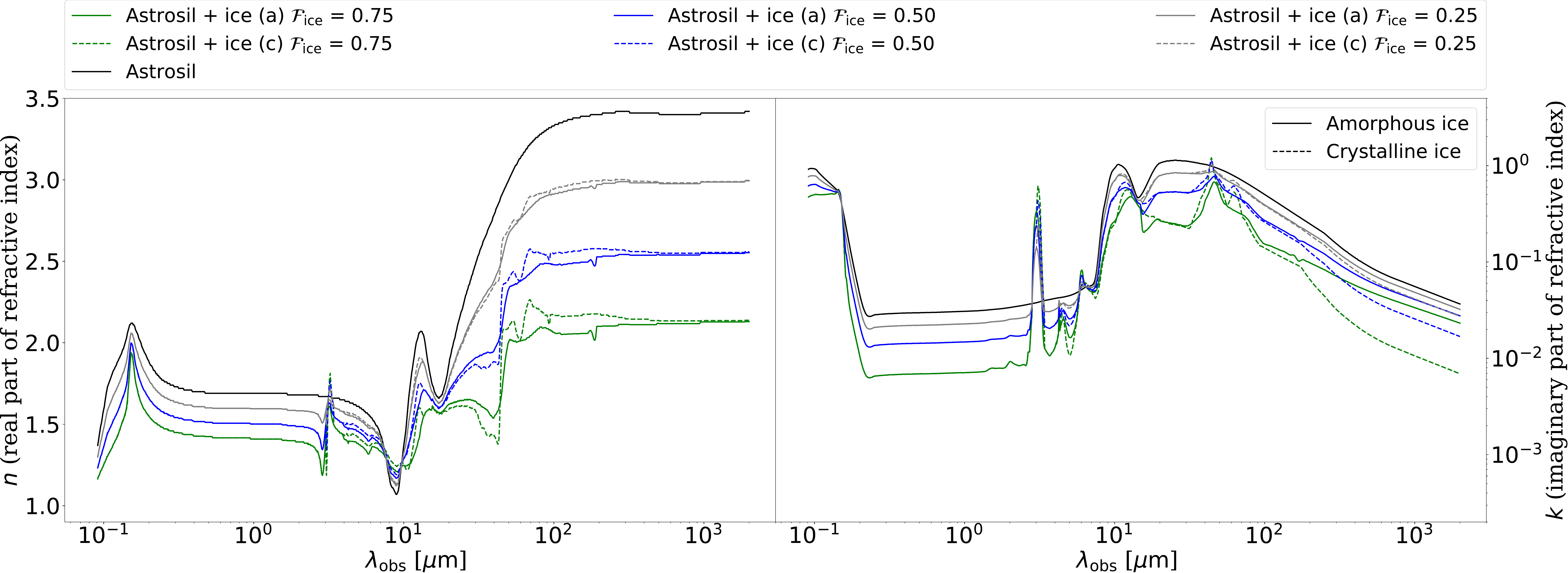}
\caption{Optical constants $n$ and $k$, i.e., the real and imaginary part of the refractive index, of icy-astrosilicate aggregate depending on the fractional ratio of ice ${\mathcal{F}}_{\rm ice}$. Ice (a), ice (c), and astrosil indicate amorphous ice (solid line), crystalline ice (dashed line), and astrosilicate, respectively.}
\label{FigVibStab}
\end{figure*}
\begin{figure*}[h!]
\centering
\includegraphics[width=18cm]{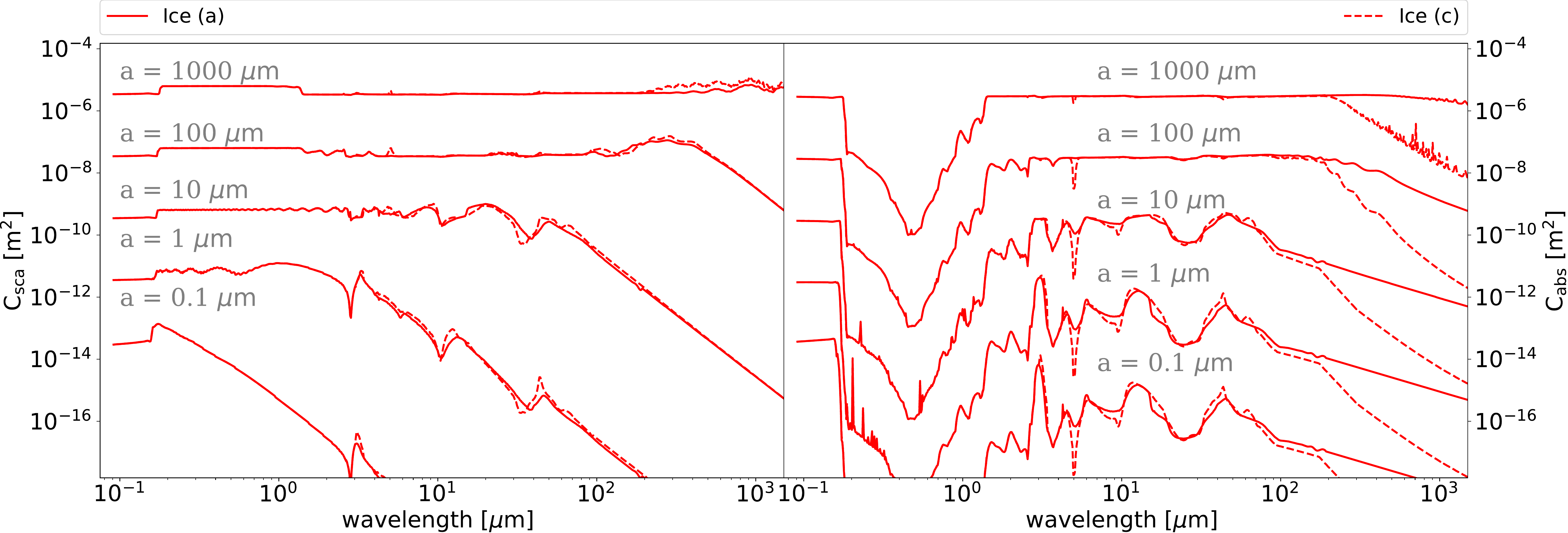}
\caption{Assumed scattering and absorption cross sections (C$_{\rm sca}$ and C$_{\rm abs}$, respectively) of amorphous ice (solid lines) and crystalline ice (dashed lines) for different grain sizes. The individual grain size is indicated in each plot. For reference to the underlying complex refractive indices, resulting from laboratory measurements, see Table 1.}
\label{FigVibStab}
\end{figure*}
\begin{figure*}[h!]
\centering
\includegraphics[width=18cm]{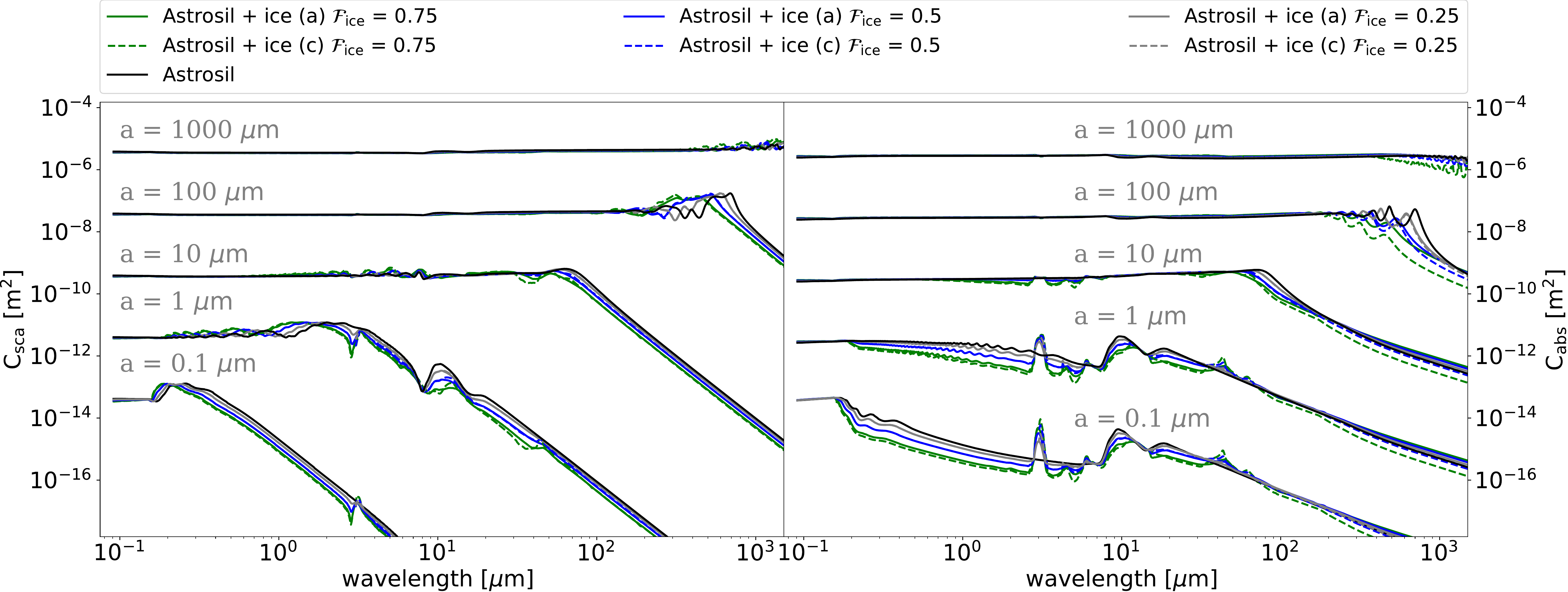}
\caption{Assumed scattering and absorption cross sections (C$_{\rm sca}$ and C$_{\rm abs}$, respectively) of ice-astrosilicate aggregates depends on the fractional ratio of ice ${\mathcal{F}}_{\rm ice}$ for different grain sizes. Ice (a), ice (c), and astrosil indicate amorphous ice (solid line), crystalline ice (dashed line), and astrosilicate, respectively. The individual grain size is indicated in each plot. For reference to the underlying complex refractive indices, resulting from laboratory measurements, see Table 1.}
\label{FigVibStab}
\end{figure*}
\section{Model description}

{\hspace{-2mm}}\noindent \textbf{Debris disk and central star}{\hspace{2mm}} In the following, the basic characteristics of our reference debris disk model are briefly summarized (see also Table 1). We considered a fiducial idealized typical debris disk system around a $\beta$ Pic-like star (A6 V main-sequence star). For the inner radius of debris disks we consider 3 au, motivated by the region close to the sublimation line of pure ice. We note that the direct observation of a spatially resolved inner radius of the debris disk system is limited by the fixed occulting spot size, for instance, 0.6" in HST observation. Alternatively, the analysis of debris disks SEDs allows constraining the inner radius as well, but is limited by uncertainties of the optical properties of dust grains. For the outer radius of debris disks we consider 150 au, motivated by spatially resolved observations of debris disks\footnote{https://www.astro.uni-jena.de/index.php/theory/catalog-of-resolved-debris-disks.html \\\noindent http://circumstellardisks.org}. \newline
\indent Optically thin debris disks are assumed to approximately cover the range of radial density profiles n(r)~$\propto$~r$^{\rm-1.0\sim-2.5}$~(\citealp{Smith}; \citealp{Artymowicz1989}; \citealp{Kalas1995}; \citealp{Pantin}; \citealp{Gor}; \citealp{Krivov2006}; \citealp{Strubbe}). We therefore consider a radial density profile of the disk of n(r) $\propto$ r$^{\rm-1.5}$ (\citealp{Krivov2006}; \citealp{Strubbe}). \newline
\indent Surveys at submillimeter (submm) wavelengths have shown that the dust mass of most debris disks typically ranges from $\sim$ 10$^{\rm-9}$ to several 10$^{\rm-7}$${\rm M}_{\odot}$ (e.g., \citealp{Greaves2005}, and references therein). We therefore consider a dust mass in debris disks of 10$^{\rm-8}$ ${\rm M}_{\odot}$. \newline

{\hspace{-2mm}}\noindent \textbf{Chemical composition of the dust}{\hspace{2mm}} The chemical composition of the dust is considered to be similar to that of the dust in the interstellar medium, mainly consisting of silicates (astronomical silicate, hereafter referred to as astrosil) and carbonaceous grains, but also of ice (\citealp{Henning}; \citealp{Draine}). To study the influence of ice parameters, we considered two basic types of ice with different physical states: amorphous and crystalline ice. We note that the sublimation temperature for both forms of ice depends on their physical state as well, for example, 100\,K for amorphous ice and 105\,K for crystalline ice (\citealp{Fraser}; \citealp{BrownR}; \citealp{BrownBolina}; \citealp{Kobayashi2011}). In addition, the sublimation temperature does not change notably for the ice-dust mixture (\citealp{Kobayashi2011}; \citealp{potapov20182}). We therefore consider 100\,K for amorphous ice dust aggregates and 105\,K for crystalline ice dust aggregates. In our model, we considered the astronomical silicates as the dust material. \newline 
\indent The chemical composition of the icy dust aggregates is defined by the fraction of the total ice volume ${\mathcal{F}}_{\rm ice}$~=~0 (corresponds to a pure astrosilicate grain), 0.25, 0.5, 0.75, and 1 (corresponding to pure ice), resulting in bulk densities from 3.5 - 0.25 g/cm$^{3}$ (\citealp{Draine}; \citealp{Kobayashi2010}). In addition, for porous ice particles we consider volume fractions of vacuum inclusions $\mathcal{P}$~=~0 (corresponding to compact ice grains), 0.25, 0.5, and 0.75, where $\mathcal{P}$~=~1~-~${V_{\rm ice}}/{V_{\rm total}}$~=~${V_{\rm vacuum}}/{V_{\rm total}}$. \newline

\indent {\hspace{-2mm}}\noindent \textbf{Inhomogeneous mixtures and fluffy structure}{\hspace{2mm}}  There are indications that interstellar and interplanetary dust grains have an inhomogeneous and fluffy structure. We applied the effective medium theory (EMT) to describe the optical properties of composite material resulting from the optical properties and relative fractions of its components. We used the code \textbf{emc} (effective medium calculator; \citealp{Ossenkopf}) to compute the effective refractive index, that is, the scattering and extinction behaviors, using rules of the effective medium approximations (i.e., Maxwell-Garnett rule) for several types of inclusions with different bulk materials. We investigated the optical properties of dust aggregates with various shapes, such as a spherical shape of ice inclusion-astrosilicate matrix particles (hereafter inclusion-matrix particles), ice mantled-astrosilicate core particles (hereafter core-mantle particles), porous ice particles, and particles with a platelet shape of ice inclusion-astrosilicate matrix (hereafter platelet-shape particles). \newline

\indent {\hspace{-2mm}}\noindent \textbf{Optical data and properties of dust}{\hspace{2mm}} In the following, the optical data of dust components we used are described briefly. In particular, these are the real and imaginary parts of the complex refractive index ($n$ and $k$, respectively) for pure amorphous ice, crystalline ice, and astrosilicate (see Figs. 1 and 2). The real and imaginary parts of the complex refractive index are fundamental parameters that determine the scattering and absorption properties of dust particles. Based on these, we derived the wavelength-dependent scattering and absorption cross sections C$_{\rm sca}$ and C$_{\rm abs}$, respectively, and corresponding scattering and absorption efficiencies~Q$_{\rm sca}$~and~Q$_{\rm abs}$~(see~Figs.~3~and~4). These optical properties also play a key role in the dynamical evolution of dust particles by modifying their lifetime thorugh the P-R effect. We find that C$_{\rm sca}$ and C$_{\rm abs}$ increase with increasing grain size at long wavelengths, regardless of their specific chemical composition. In addition, the strength of various characteristic features decreases with increasing grain size. \newline
\indent For the optical data of crystalline ice in the visual (VIS) to near-IR range, that is, at 0.1~$\mu\rm{m}$ to 2~$\mu\rm{m}$, we used the data from \citet{Li}. In the near-IR to far-IR range, that is, at 2~$\mu\rm{m}$ to 94~$\mu\rm{m}$, we extrapolated and incorporated the optical constants that were obtained from transmission spectra of pure crystalline ice by \citet{Potapov} and \citet{Curtis}. In the far-IR to submm range, i.e., at 94~$\mu\rm{m}$ to 1000~$\mu\rm{m}$, we extrapolate and derive a new set of data of crystalline ice from \citet{Reinert}, \citet{Haessner}, and \citet{Warren}. We note that the slope of the imaginary part of refractive index $k$ sensitively depends on the temperature with increasing wavelength $\lambda$ from 175~$\pm$~6~$\mu\rm{m}$, i.e., getting steeper at a lower temperature (\citealp{Mishima}; \citealp{Reinert}; \citealp{Haessner}; see Fig. 1). \newline 
\indent For the optical data of amorphous ice in the VIS/near-IR to far-IR/sub-mm range, i.e., at 0.1~$\mu\rm{m}$ to 2~$\mu\rm{m}$ and at 200~$\mu\rm{m}$ to 1000~$\mu\rm{m}$, we use the optical constants from \citet{Li}, which are based on the data from \citealp{Hudgins} with power-law extrapolation in the sub-mm range. In the near to far-IR range, i.e., at 2~$\mu\rm{m}$ to 200~$\mu\rm{m}$, we use the transmission spectra of pure amorphous ice from \citet{Potapov} and \citet{Curtis}). For the optical data of silicate in the optical to submm range, that is, at 0.1 to 1000~$\mu\rm{m}$, we used the data from \citet{Draine}. \newline\newline
\indent {\hspace{-2mm}}\noindent \textbf{Grain size distribution}{\hspace{2mm}} In a steady-state grain size distribution, $n(a)$ follows the power-law distribution $n(a)$ $\propto$ $a^{\rm - 3.5}$ (\citealp{Dohnanyi}), which represents an approximation for grains around the blow-out size up to planetesimal size, resulting from a collisional cascade (e.g., \citealp{Thebault2003}; \citealp{Krivov2006}; \citealp{Thebault2007}; \citealp{Krivov2008}). In summary, we considered grain sizes from 0.1 to 1000 $\rm{\mu m}$ with the above steady-state grain size distribution. We note that the nongravitational forces acting on grains particularly in the range of tens to hundreds of micrometers in diameter may further modify the size distribution (\citealp{Krivov2000}; \citealp{Krivov2006}; \citealp{Plavchan}; \citealp{Loehne2017}; \citealp{Kim}). However, this effect has not been taken into account in the current study. \newline

\indent {\hspace{-2mm}}\noindent \textbf{Ice destruction mechanisms}{\hspace{2mm}} Using the approach and results from \citet{Grigorieva}, we considered ice destruction mechanisms through UV photosputtering and collisions. In particular, they found that only~$\textgreater$~5~mm grains can retain their ice~at~$\sim$~80~au. We therefore considered ice grains with radii~5~mm and~80~au as the smallest dust grain size and inner radius of debris disks in case of UV photosputtering in our model. However, we considered ice grains with radii~20~$\mu\rm{m}$ and~40~au as the smallest dust grain size and inner radius of debris disks in case of UV photosputtering and collisions. This is because \citet{Grigorieva} showed that the collisional activity increases the abundance of smaller ice grains in the inner region of debris disks.\newline
\indent In addition, sublimation of ice was considered. The sublimation radius was derived from the sublimation temperature of each considered dust or ice species and the corresponding radial temperature distribution. The latter was calculated on the basis of the optical properties of the dust or ice (which are in turn a function of the complex refractive index, the shape, and internal structure of the considered dust or ice species). At the inner part of the debris disk system, where only astrosilicate is present as a result of ice sublimation, we applied the optical properties of porous astrosilicate with the same shape of dust aggregate. Thus, the chemical composition of astrosilicate is defined by the fraction of the vacuum inclusions $\mathcal{P}$~=~${V_{\rm vacuum}}/{V_{\rm total}}$, which is equal to the fractional ratio of sublimated ice.\newline\newline

\indent {\hspace{-2mm}}\noindent \textbf{The simulation of observables of debris disks}{\hspace{2mm}} We used a newly developed software tool called debris disks around main-sequence stars (DMS; \citealp{Kim}), which is optimized for simulating observables of debris disks, or in other words, optically thin systems. In particular, it allows us to simulate scattered light and thermal dust reemission images, the continuum spectral energy distribution (SED), and scattered light polarization images. The optical properties of the dust grains were computed using the tool \textbf{miex} (\citealp{Wolf}). The stellar photospheric emission corresponding to the chosen stellar parameters was taken from the \textbf{PHOENIX/NextGen} database (\citealp{Hauschildt}). \newline
\indent This study is the simulation approach to implement the temperature-dependent optical data (\citealp{Omont}; \citealp{Robinson}; see Fig.~1) of crystalline ice laboratory data from \citet{Reinert} and \citet{Haessner}. In the first step, we calculated the radial temperature distribution. For this purpose, we applied the optical data measured at a temperature of 55K, that is, the optical data corresponding to a median temperature. However, to calculate observable quantities (SEDs, images, and polarization), we then applied the optical data corresponding to the temperature distribution calculated before.

\section{Results}

\noindent In the following, we discuss and analyze the effects of dust parameters and various ice destruction mechanism that were discussed in Section 2 on the resulting SED (Section 4.1) and spatially resolved images (Section 4.2). In addition, we predict the corresponding ice survival line of debris disks (Section 4.3). We finally focus on constraining the detectability of ice in debris disk systems with future observation by the JWST/NIRCam and SPICA/SAFARI (Section 4.4).
\begin{figure}[h!]
\centering
\includegraphics[width=9cm]{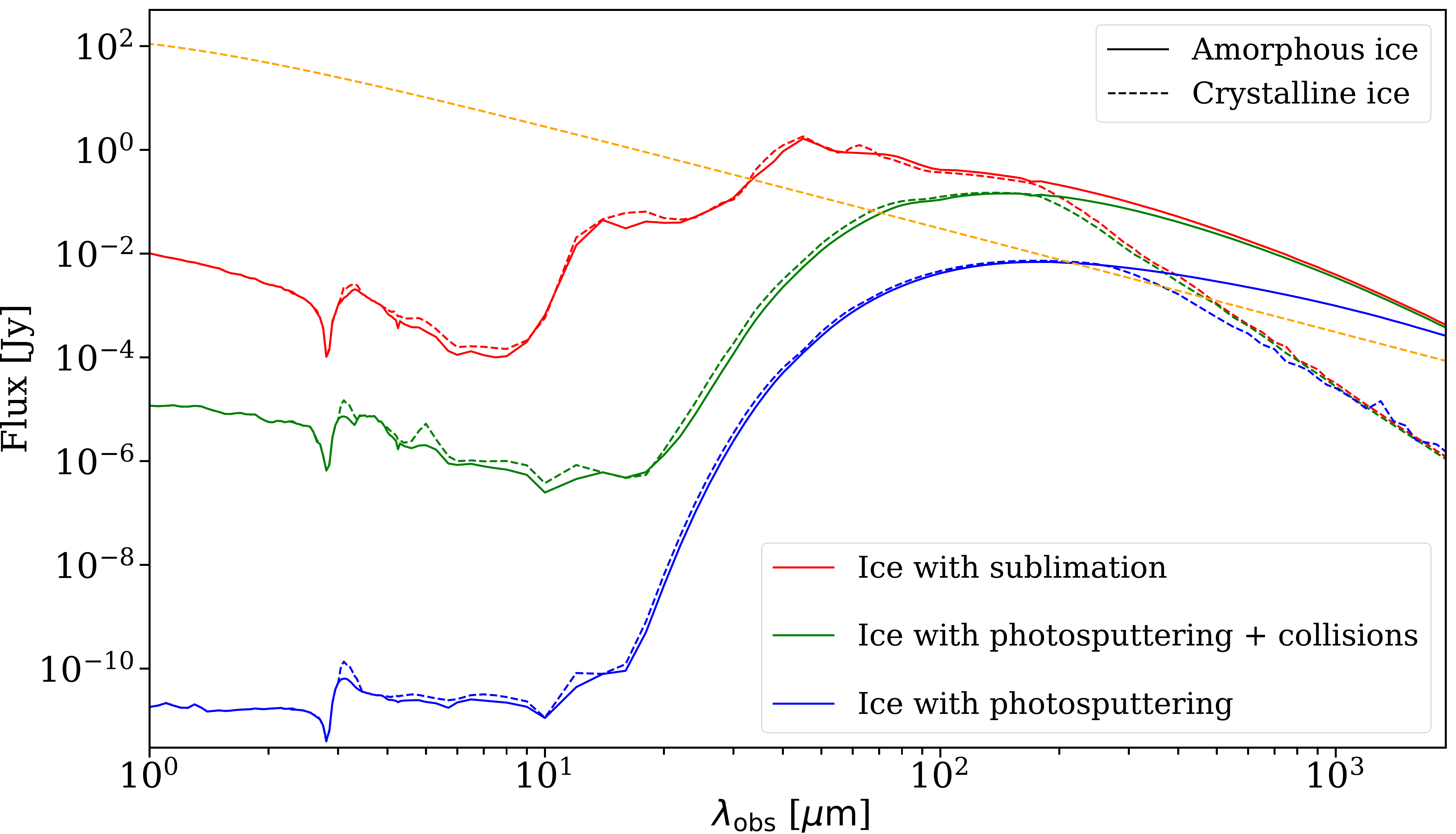}
\caption{Effect of ice destruction mechanisms on the resulting SED. UV photosputtering and mutual collisions are considered in addition to ice sublimation. The dashed yellow line represents the photospheric emission of the central star. The solid and dashed lines indicate amorphous ice and crystalline ice with ${\mathcal{F}}_{\rm ice}$~=~1, respectively.}
\label{FigVibStab}
\end{figure}
\begin{figure}[h!]
\centering
\includegraphics[width=9cm]{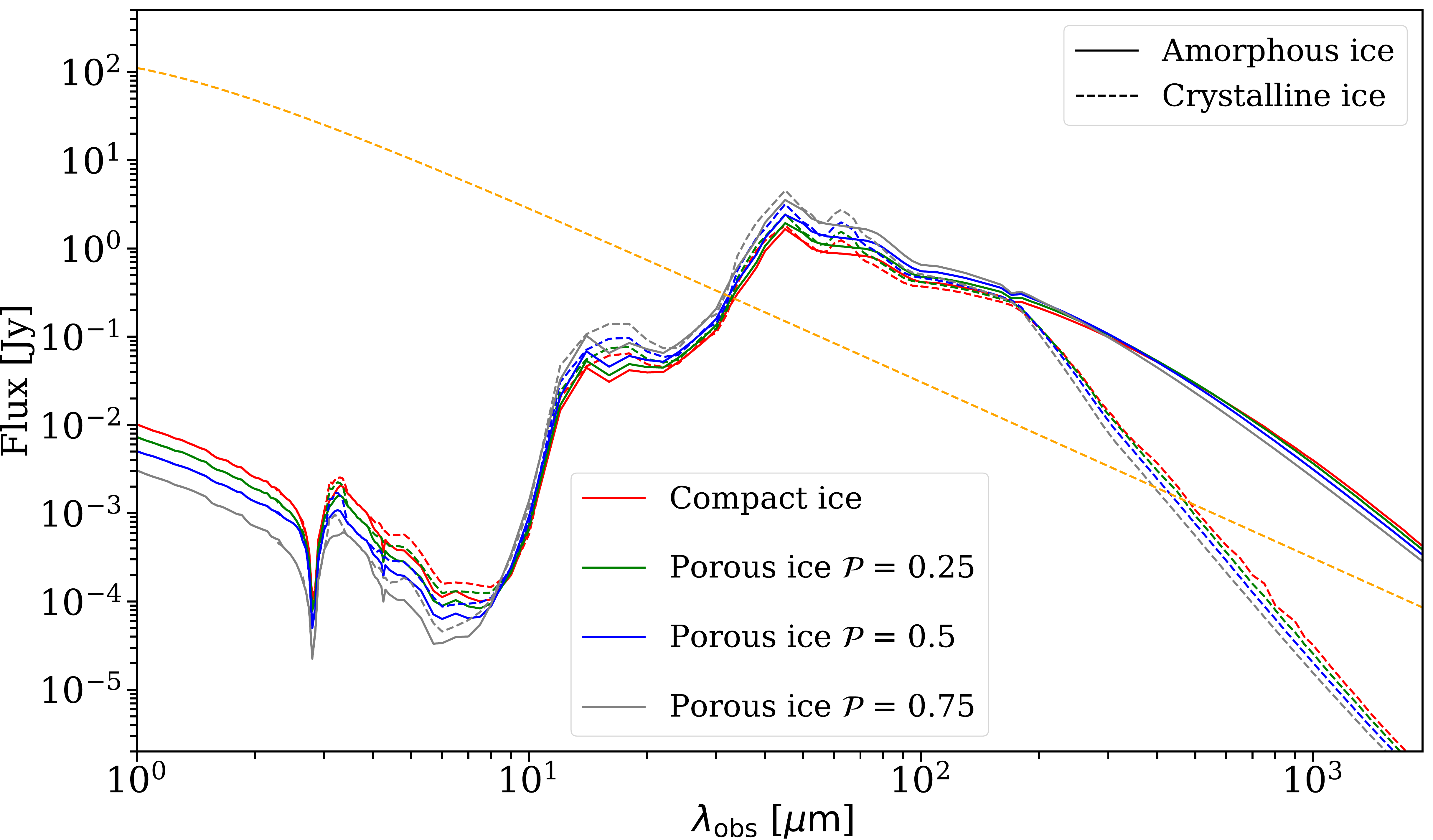}
\caption{Effect of ice porosity on the resulting SED. $\mathcal{P}$ indicates the porosity of ice grains. The dashed yellow line represents the photospheric emission of the central star. The solid and dashed lines indicate amorphous ice and crystalline ice, respectively.}
\label{FigVibStab}
\end{figure}
\begin{figure*}[h!]
\centering
\includegraphics[width=18cm]{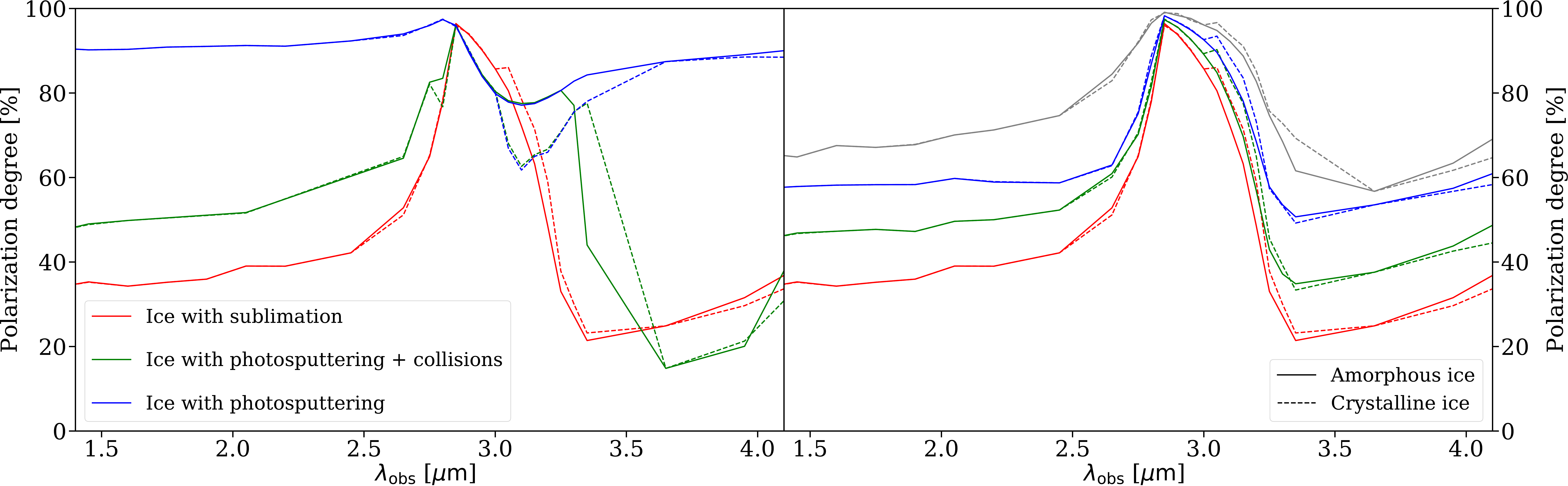}
\caption{Effect of the ice destruction mechanism (left) and porosity (right) on the wavelength-dependent polarization degree at near-IR to mid-IR wavelengths. Ice (a) and ice (c) indicate amorphous (solid line) and crystalline ice (dashed line) with ${\mathcal{F}}_{\rm ice}$~=~1, respectively.}
\label{FigVibStab}
\end{figure*}

\subsection{Spectral energy distribution}
\noindent We first investigate the influence of individual dust parameters, that is, the fractional ratio of ice ${\mathcal{F}}_{\rm ice}$, the different shapes of the aggregates, and porosity of ice, on the resulting SED. In addition, we investigate the ice grain survival by quantitatively exploring the role of UV photosputtering and mutual collisions in addition to the sublimation mechanism. We focus on IR to submm wavelengths because the offset from the stellar photospheric SED is largest in this wavelength region. The specific absorption or scattering features are reflected in the corresponding features of the resulting SED (see Figs. 3 and 4).\newline

{\hspace{-2mm}}\noindent \textbf{4.1.1. Pure ice} {\hspace{2mm}} In Fig. 5 we show the SED of debris disks composed of pure ice (${\mathcal{F}}_{\rm ice}$~=~1) considering three different mechanisms of ice destruction: UV photosputtering, mutual collisions and sublimation. First, we find that UV photosputtering is responsible for the destruction of small ice grains; thus, scattered radiation from the ice is significantly decreased in the near-IR to mid-IR wavelength range. Consequently, UV photosputtering reduces the flux in near-IR to mid-IR wavelengths by~$\text{about}$~eight orders of magnitude compared to the case in which only sublimation is considered. Even if the collisional activity is taken into account, the flux is still decreased by~$\text{about}$~2 orders of magnitude. Furthermore, we find that UV photosputtering (and collisions) are responsible for the strength of the ice features, for instance, the shallow 3~$\mu\rm{m}$ feature, which is in good agreement with previous studies (\citealp{Kamp}; \citealp{Honda16}). \newline 
\indent UV photosputtering (and collisions) significantly contribute to the erosion of the decreased flux far beyond the sublimation-imposed ice survival line as well. This is reflected by the decreased flux at far-IR wavelengths. Consequently, this results in a weakened ice features around 20 - 30~$\mu\rm{m}$ and the shift of the location of the maximum of the dust reemission flux from $\leq$ 40~$\mu\rm{m}$ for sublimation toward $\leq$ 200~$\mu\rm{m}$ ($\leq$ 100~$\mu\rm{m}$ for collisions). We also find a decrease of the peak flux in the case of UV photosputtering (and collisions). In contrast to the previous finding of a shallow 3~$\mu\rm{m}$ feature, other ice specific features, for instance, the 44~$\mu\rm{m}$ and 62~$\mu\rm{m}$ of crystalline ice and 44~$\mu\rm{m}$ of amorphous ice, disappear when UV photosputtering (and collisions) are considered. \newline 
\indent In the case of the largest grains, UV photosputtering (and collisions) can no longer contribute efficiently to destruction and erosion processes. Therefore, the effect of UV photosputtering (and collision) of submm wavelengths is weaker, or in other words, less pronounced. Consequently, the SED in this wavelength range has a similar spectral index, regardless of the destructive mechanisms that make it hardly possible to constrain the mechanisms of ice destruction from the analysis of the SED alone. In addition, because the absorption coefficient, for example, C$_{\rm abs}$ of crystalline ice, is lower than that of amorphous ice (see Fig. 3), their spectral index also becomes significantly lower than that of amorphous ice. \newline

\noindent In Fig. 6 the SED of debris disks is shown as a function of ice grain porosity, that is, for $\mathcal{P}$~=~0 (pure compact ice), 0.25, 0.5, and 0.75. Similar to \citet{Brunngraeber}, we find only a weak influence of the porosity on the resulting SED. Highly porous ice grains show a higher peak flux in the 10~$\mu\rm{m}$ to 80~$\mu\rm{m}$ range, but a lower peak flux at shorter or longer wavelengths. This is because the different contribution from individual grain size in each wavelength range, for instance, highly porous grains with radii of about tens of a micron have a high absorption cross-section at far-IR wavelengths. \newline

{\hspace{-2mm}}\noindent \textbf{Scattered-light polarization:} While spectropolarimetric observations have shown enhanced polarization levels in the 3\,$\mu\rm{m}$ ice band for molecular clouds where the increased absorption efficiency is of importance (\citealp{Hough}; \citealp{Greenberg}; reference therein), it remains to be shown whether the feature is also of importance in scattered light. This is relevant for optically thin debris disks.\newline 
\indent We find that the various mechanisms of ice destruction significantly affect the polarimetric signal. UV photosputtering in particular results in a very high polarization degree at near-IR to mid-IR wavelengths (see the blue line in the left plot of Fig. 7), while it is decreased when collisions or sublimation were taken into account (see the green and red lines in the left plot of Fig. 7). Overall, the polarization degree is higher for smaller grains at near-IR wavelengths, with a maximum around the 3\,$\mu\rm{m}$ ice feature. In addition, we find that highly porous ice grains tend to produce high polarization degrees at near-IR wavelengths (see the right plot of Fig. 7). \newline

{\hspace{-2mm}}\noindent \textbf{4.1.2. Icy-astrosilicate dust aggregates} {\hspace{2mm}} In Fig. 8 we illustrate that the SED of debris disks that are composed of icy dust aggregates depends on the fractional ratio of ice ${\mathcal{F}}_{\rm ice}$. Astrosilicate clearly dominates the emissivity (see. Fig. 4) even if its relative fraction is as low as 10 \%. It is therefore expected that the SED of a mixture of ice and astrosilicate is similar to that of astrosilicate alone (Fig. 8). In addition, the sublimation temperature of astrosilicate is significantly higher than the two different physical states ice, that is, crystalline ice and amorphous ice. Consequently, they show a clear difference to the major fraction of warmer dust grains around near-IR to mid-IR on the resulting SED. This results in an increase of the SED in the corresponding wavelength range, which shifts the flux maximum on the SED at wavelengths~$~\text{of about}$~10~$~\mu\rm{m}$. \newline
\indent Finally, we investigate the effect of the fractional ratio of ice ${\mathcal{F}}_{\rm ice}$ on the observation of ice features. The prominent 3\,$\mu\rm{m}$ ice feature can be found even in ice-poor aggregates, that is, for a lower fractional ratio of ice ${\mathcal{F}}_{\rm ice}$. In contrast, the ice features at 44~$\mu\rm{m}$ and 62~$\mu\rm{m}$ remain only in ice-rich aggregates, that is, in grains with higher fractional ratios of ice ${\mathcal{F}}_{\rm ice}$. Interestingly, the usually very prominent 10~$\mu\rm{m}$ astrosilicate feature disappears for most of the considered icy dust mixtures. \newline
 
\begin{figure}
\centering
\includegraphics[width=9cm]{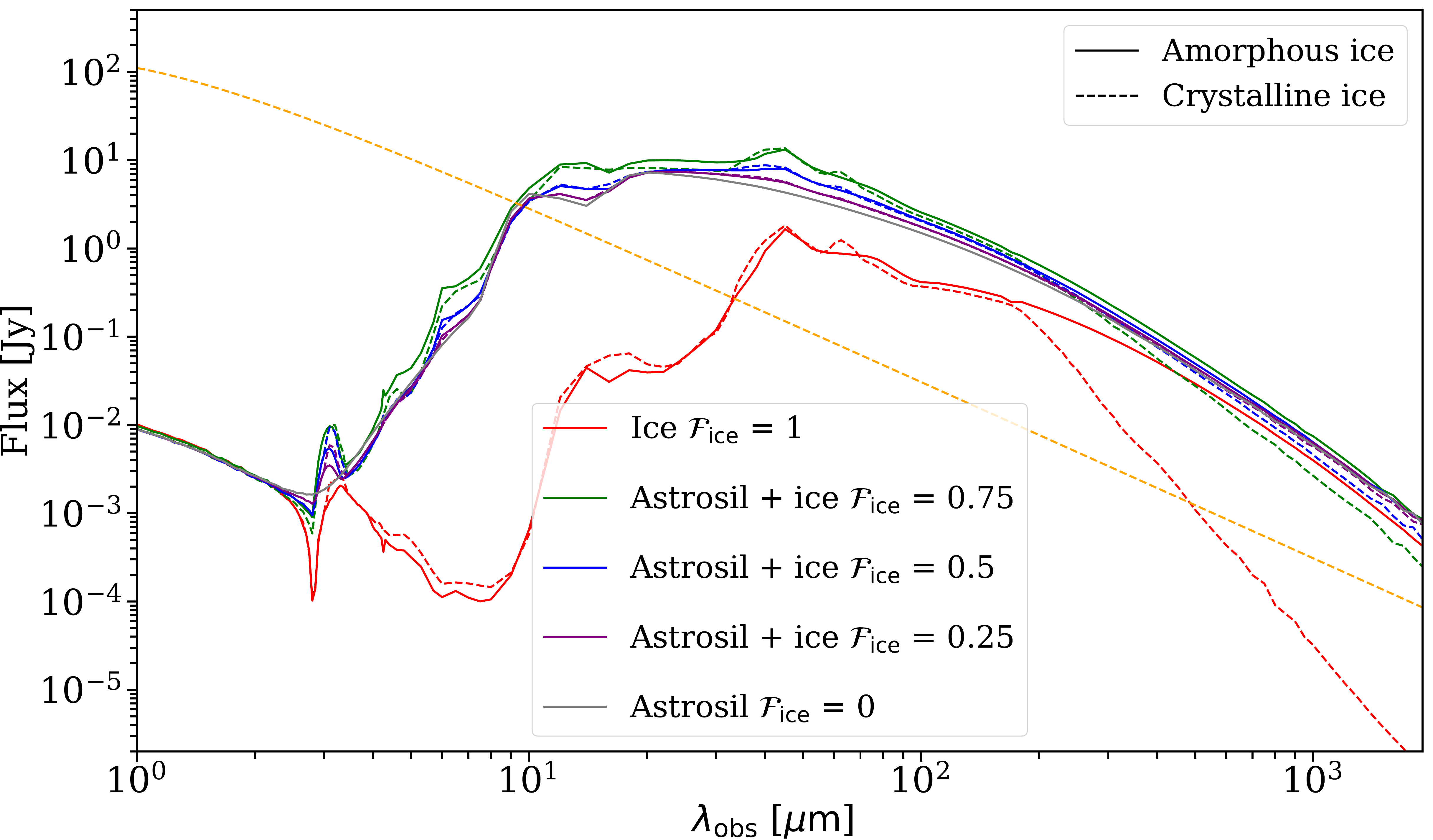}
\caption{Effect of the fractional ratio of ice ${\mathcal{F}}_{\rm ice}$ on the resulting SED. The dashed yellow line represents the photospheric emission of the central star. Ice (a), ice (c), and astrosil indicate amorphous ice (solid line), crystalline ice (dashed line), and astrosilicate, respectively.}
\label{FigVibStab}
\end{figure}
\begin{figure}
\centering
\includegraphics[width=9cm]{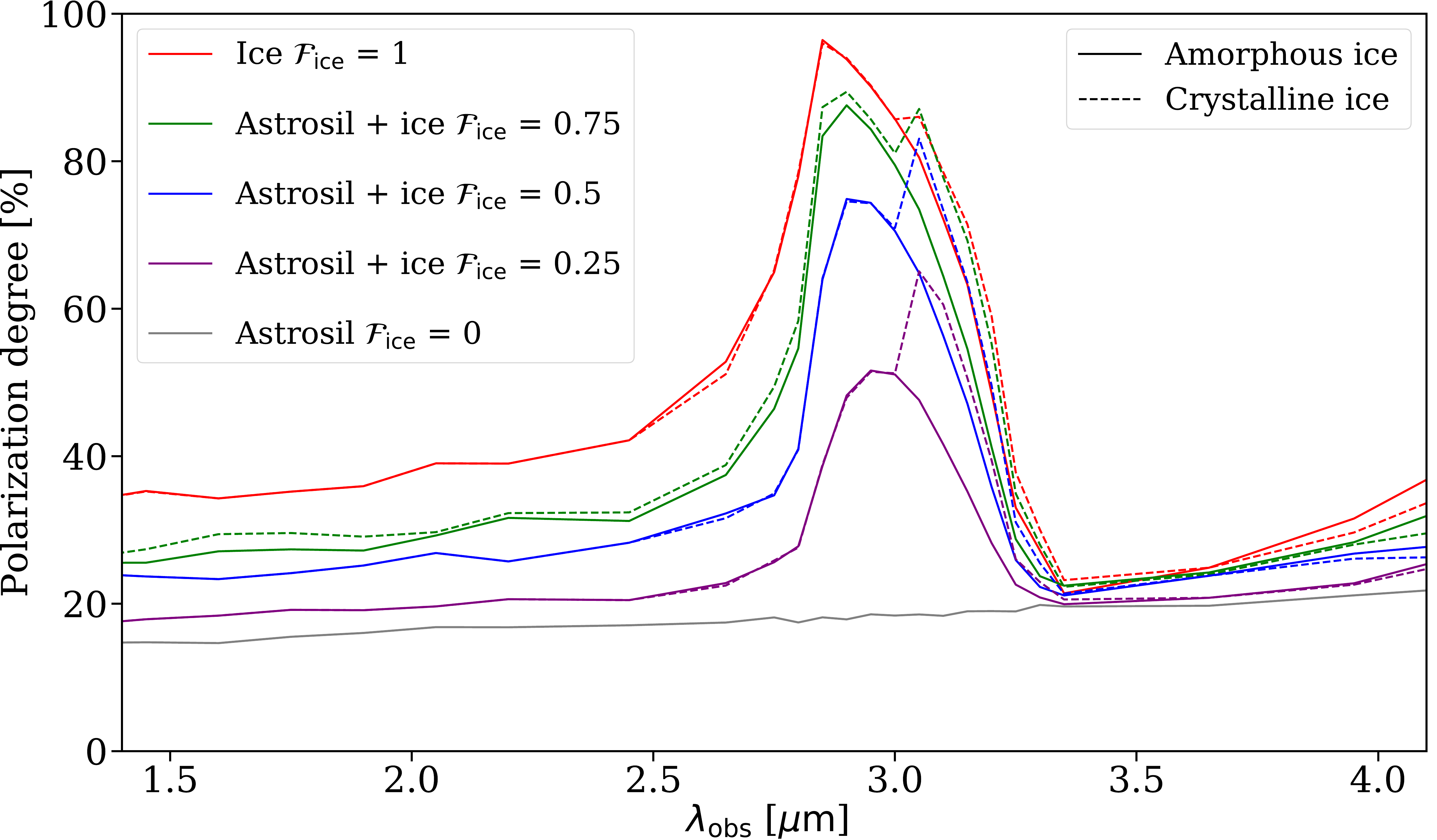}
\caption{Effect of the fractional ratio of ice ${\mathcal{F}}_{\rm ice}$ on the wavelength-dependent polarization degree at near-IR to mid-IR wavelengths. Ice (a), ice (c), and astrosil indicate amorphous ice (solid line), crystalline ice (dashed line), and astrosilicate, respectively.} 
\label{FigVibStab}
\end{figure}
\begin{figure}
\centering
\includegraphics[width=9cm]{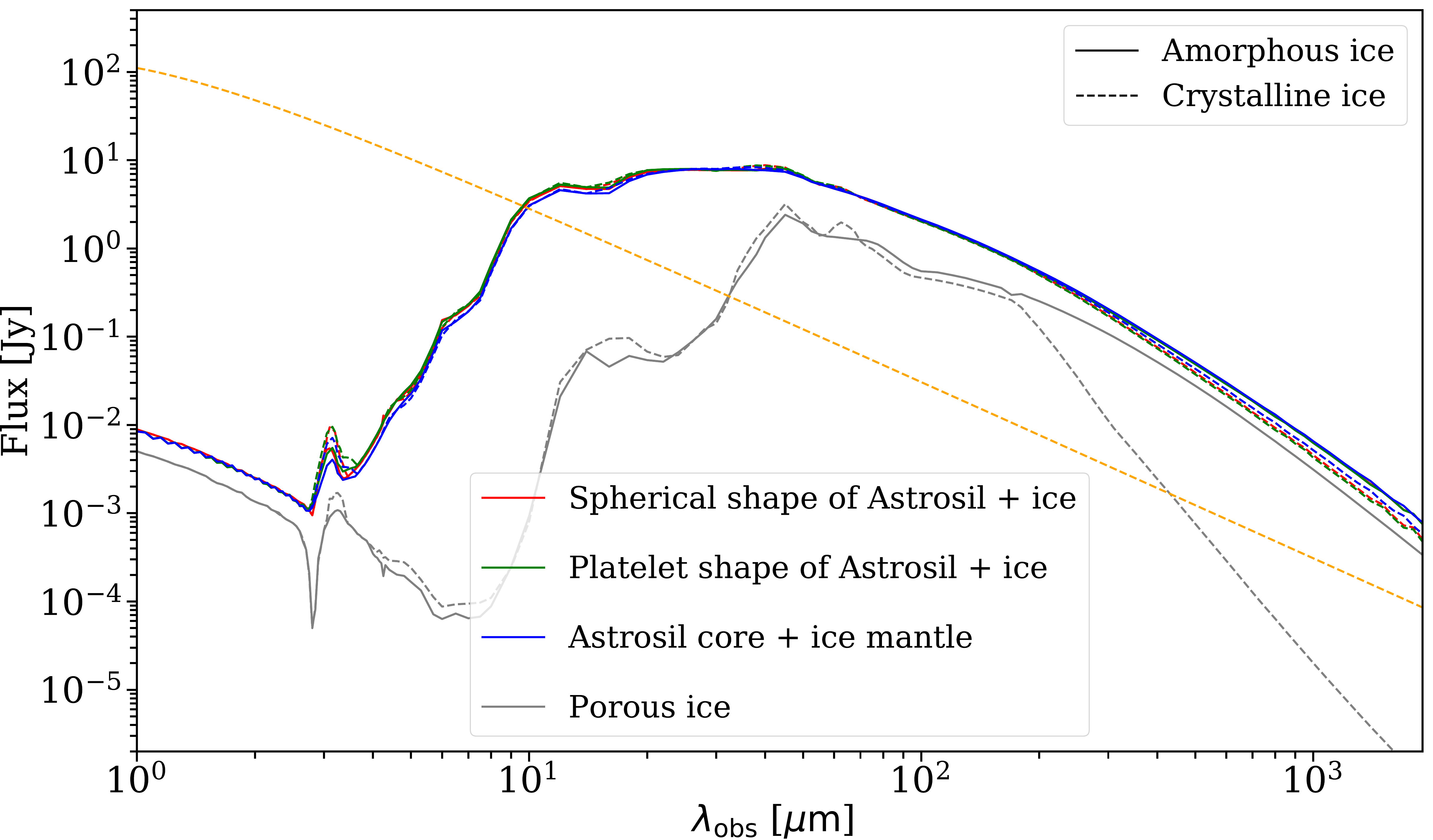}
\caption{Effect of the shape of dust aggregates on the resulting SED. Inclusion-matrix particles and core-mantle particles with spherical shape, inclusion-matrix particles with platelet shapes, and porous ice, with the same fractional ratio of ice (${\mathcal{F}}_{\rm ice}$~=~0.5) are considered. The dashed yellow line represents the photospheric emission of the central star. Ice (a), ice (c), and astrosil indicate amorphous ice (solid line), crystalline ice (dashed line), and astrosilicate, respectively.}
\label{FigVibStab}
\end{figure}
\begin{figure*}
\centering
\includegraphics[width=18cm]{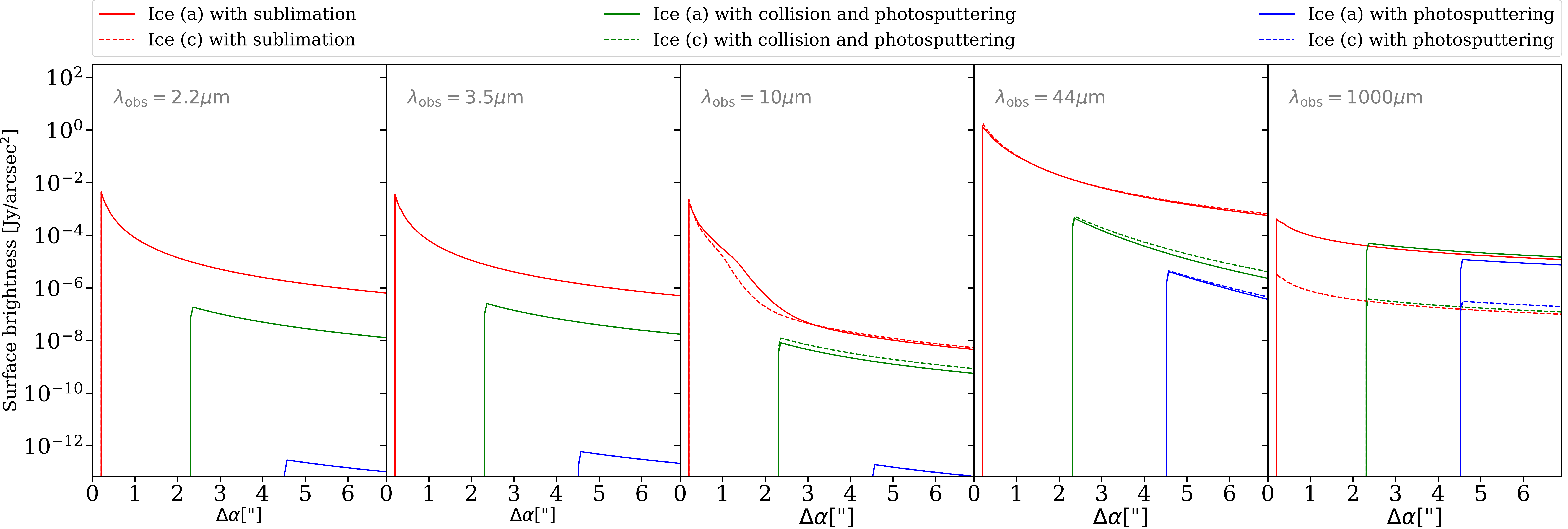}
\caption{Effect of the ice destruction mechanisms on the radial surface brightness profile at $\lambda_{\rm obs}$~=~2.2\,$\mu\rm m$, 3.5\,$\mu\rm m$, 10\,$\mu\rm m$, 44\,$\mu\rm m$, and 1000\,$\mu\rm m$.}
\label{FigVibStab}
\end{figure*}
\begin{figure*}
\centering
\includegraphics[width=18cm]{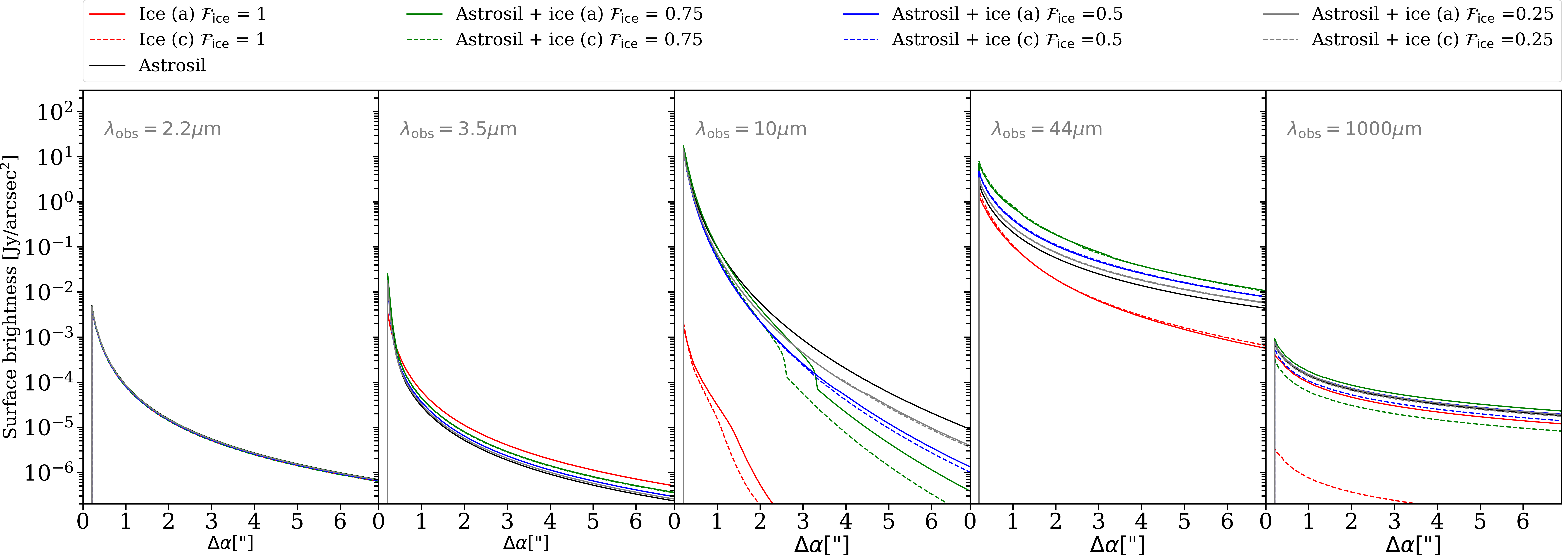}
\caption{Effect of the fractional ratio of ice ${\mathcal{F}}_{\rm ice}$ on the radial surface brightness profile at $\lambda_{\rm obs}$~=~2.2\,$\mu\rm m$, 3.5\,$\mu\rm m$, 10\,$\mu\rm m$, 44\,$\mu\rm m$, and 1000\,$\mu\rm m$.}
\label{FigVibStab}
\end{figure*}
\begin{figure*}
\centering
\includegraphics[width=18cm]{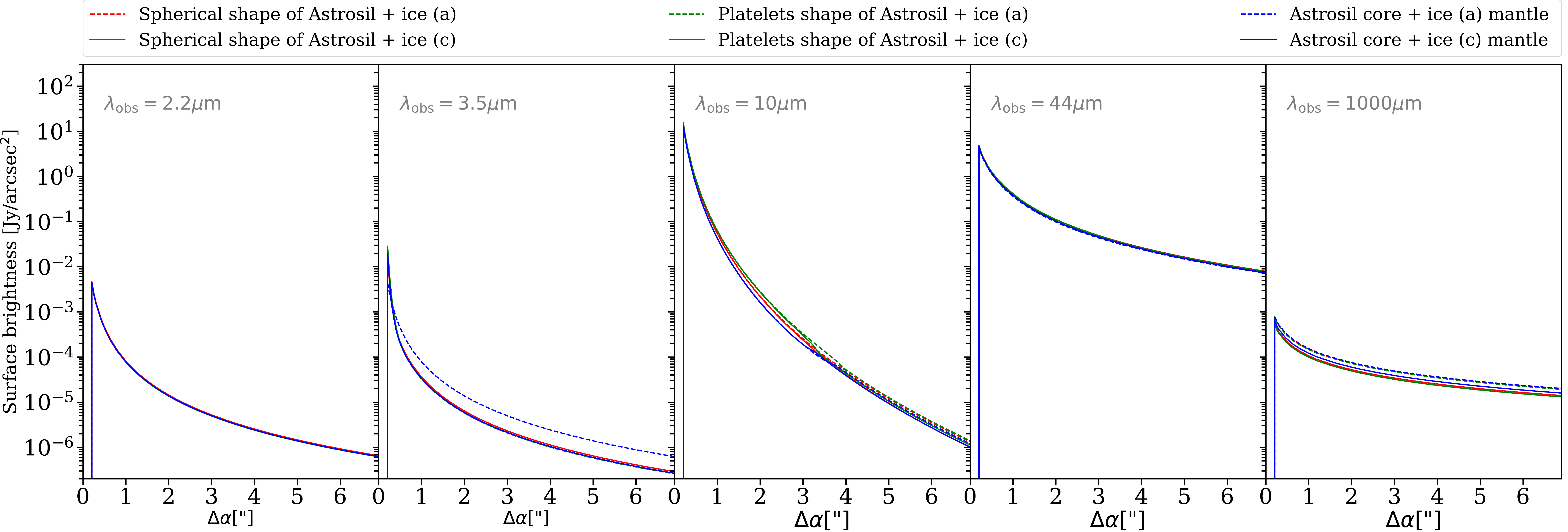}
\caption{Effect of the shape of aggregates (with same ${\mathcal{F}}_{\rm ice}$~=~0.5) on the radial surface brightness profile at $\lambda_{\rm obs}$~=~2.2\,$\mu\rm m$, 3.5\,$\mu\rm m$, 10\,$\mu\rm m$, 44\,$\mu\rm m$, and 1000\,$\mu\rm m$.}
\label{FigVibStab}
\end{figure*}

{\hspace{-2mm}}\noindent \textbf{Scattered-light polarization:} Fig. 9 shows the wavelength-dependent polarization degree at near-IR to mid-IR wavelengths as a function of the fractional ratio of ice ${\mathcal{F}}_{\rm ice}$. We find that dust grains with a higher ice fractional ratio of water ${\mathcal{F}}_{\rm ice}$ tend to produce higher polarization degrees, for example, a polarization degree of pure ice$\text{}$ of about 85 $\%$, at 3\,$\mu\rm{m}$. However, the polarization degree is lower for pure astrosilicate grain. In the submm range, this effect is less pronounced. Consequently, the measurement of the wavelength-dependent polarization degree potentially allows constraining the composition, that is, the fractional ratio of ice ${\mathcal{F}}_{\rm ice}$ or porosity of ice $\mathcal{P}$, of icy dust grains (see Fig. 7 and 9). The polarization was calculated for a scattering angle of 90$^{\circ}$. This case corresponds to spatially resolved polarization observations~of~debris~disks~seen~in~face-on~orientation. \newline 
\begin{figure*}[h!]
\centering
\includegraphics[width=18cm]{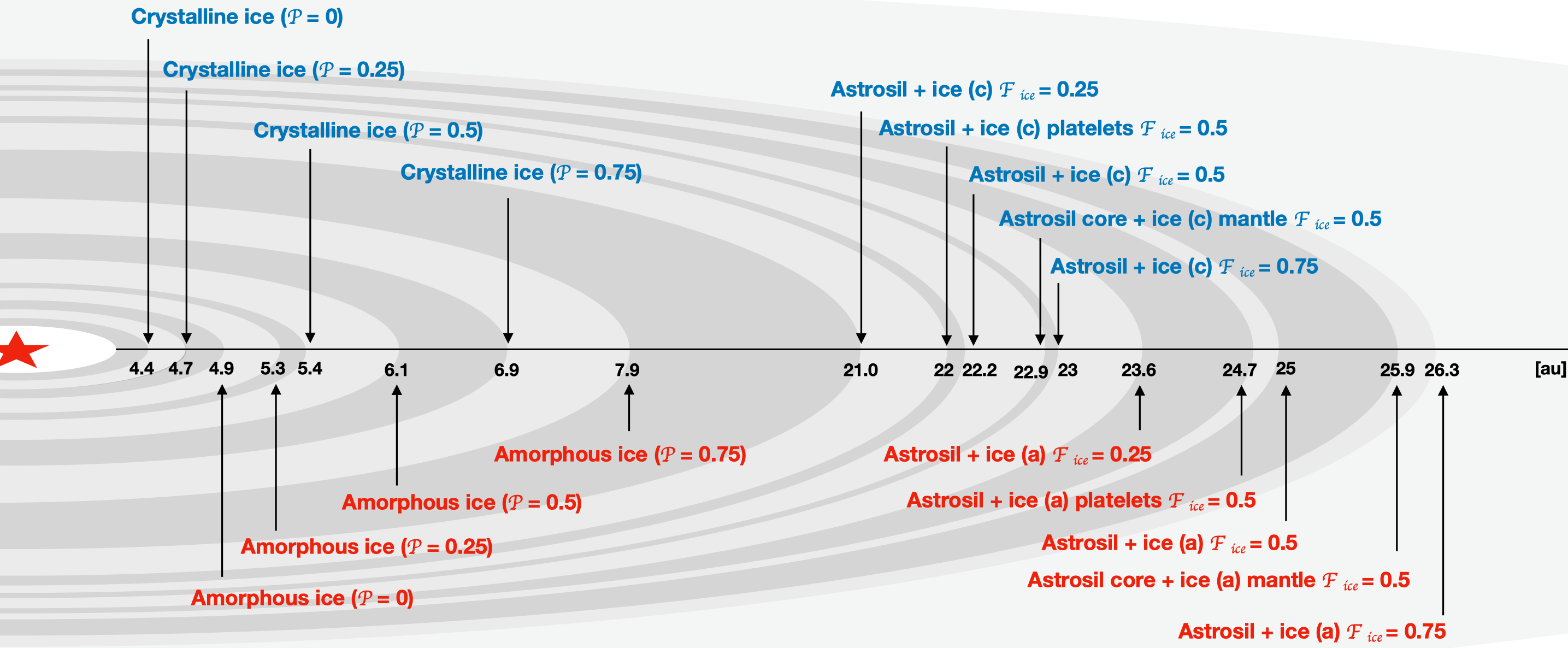}
\caption{Prediction of the location of the ice survival line for grains of blowout size in the considered $\beta$ Pic-like debris disk system. We show the dependence on the chemical component, shape, and physical state (amorphous vs. crystalline) of the icy dust aggregates.}
\label{FigVibStab}
\end{figure*}
\begin{figure}[h!]
\centering
\includegraphics[width=8.7cm]{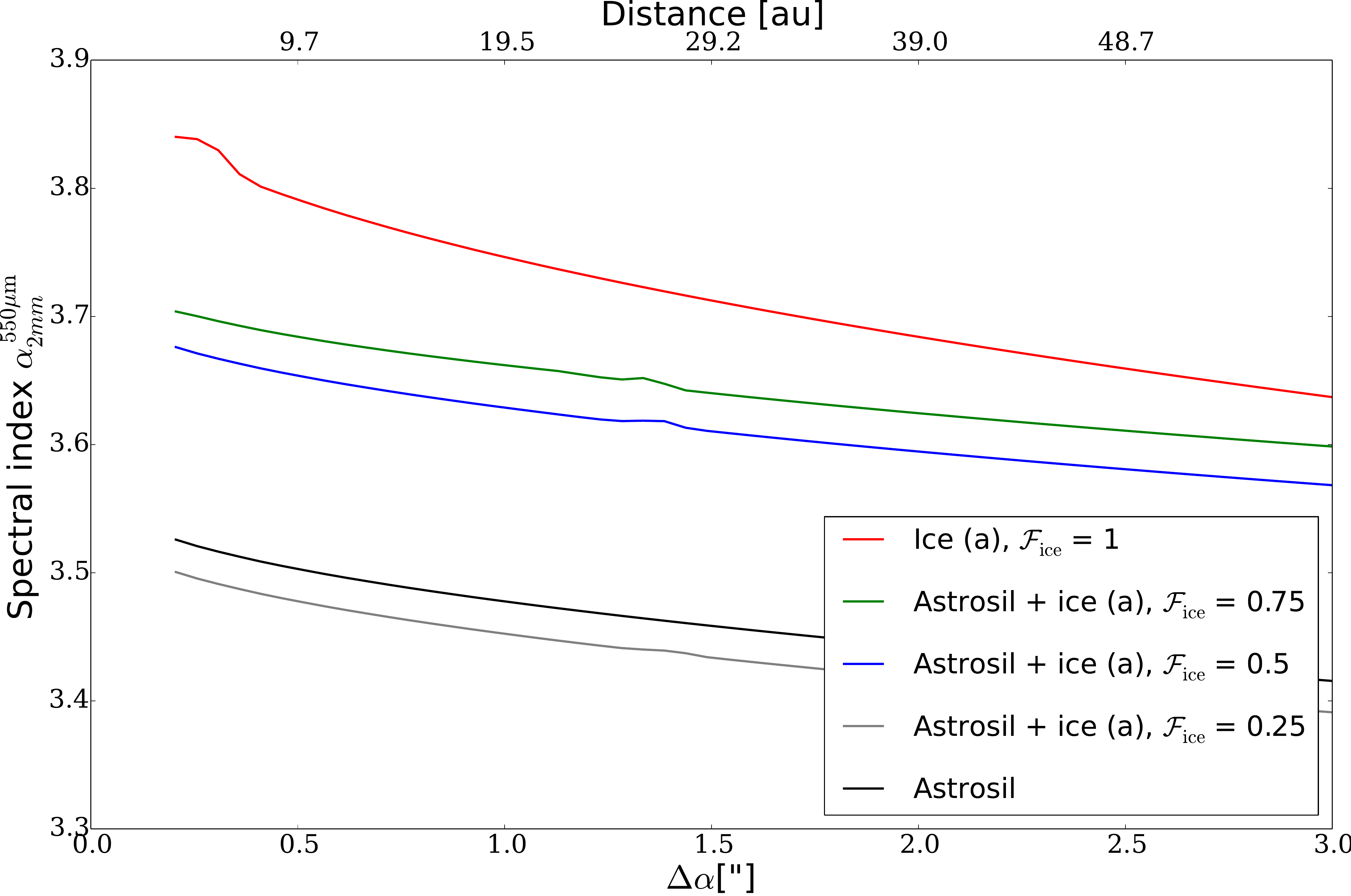}
\caption{Radial cut of spectral index $\alpha^{550\mu\rm{m}}_{2\rm{mm}}$ maps for models using the different fractional ratio of ice ${\mathcal{F}}_{\rm ice}$ (0 to 1). Ice (a) and astrosil indicate amorphous ice and astrosilicate, respectively.}
\label{FigVibStab}
\end{figure}
\begin{figure*}[h!]
\centering
\includegraphics[width=18cm]{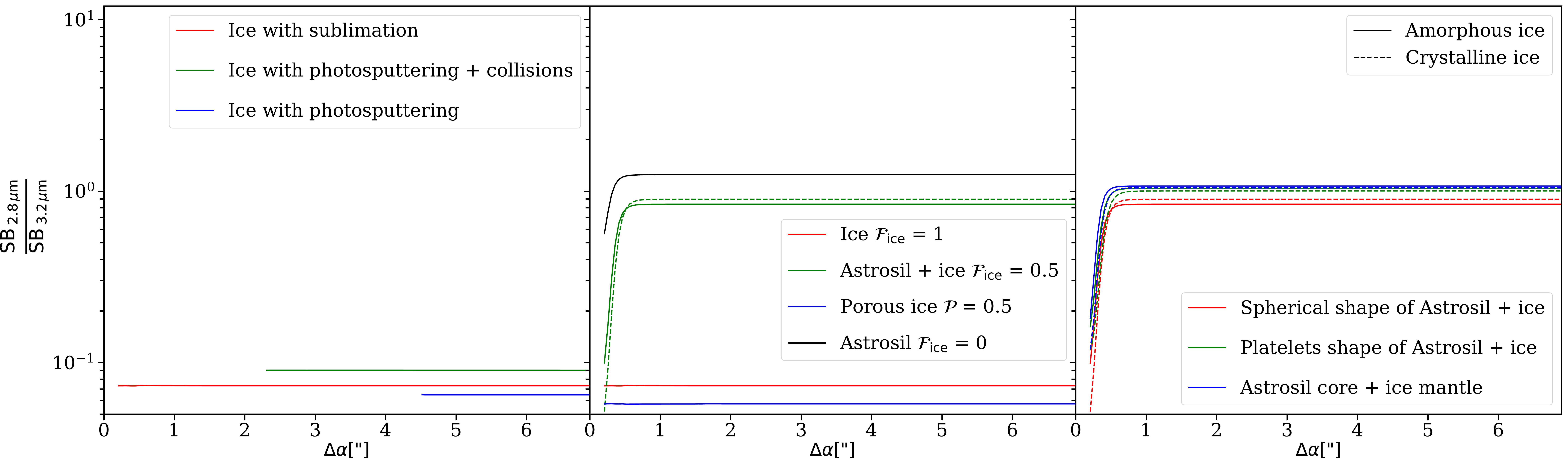}
\caption{Ratio between the surface brightness (SB) of debris disks assuming different mechanisms of ice destruction, chemical components, and shapes of icy-astrosilicate mixture at 2.8\,$\mu\rm m$ (i.e., outside of the 3\,$\mu\rm m$ ice feature) and 3.2\,$\mu\rm m$ (i.e., inside of the 3\,$\mu\rm m$ ice feature). The solid line and the dashed line indicate amorphous ice and crystalline ice, respectively.}
\label{FigVibStab}
\end{figure*}
\begin{figure*}[h!]
\centering
\includegraphics[width=18cm]{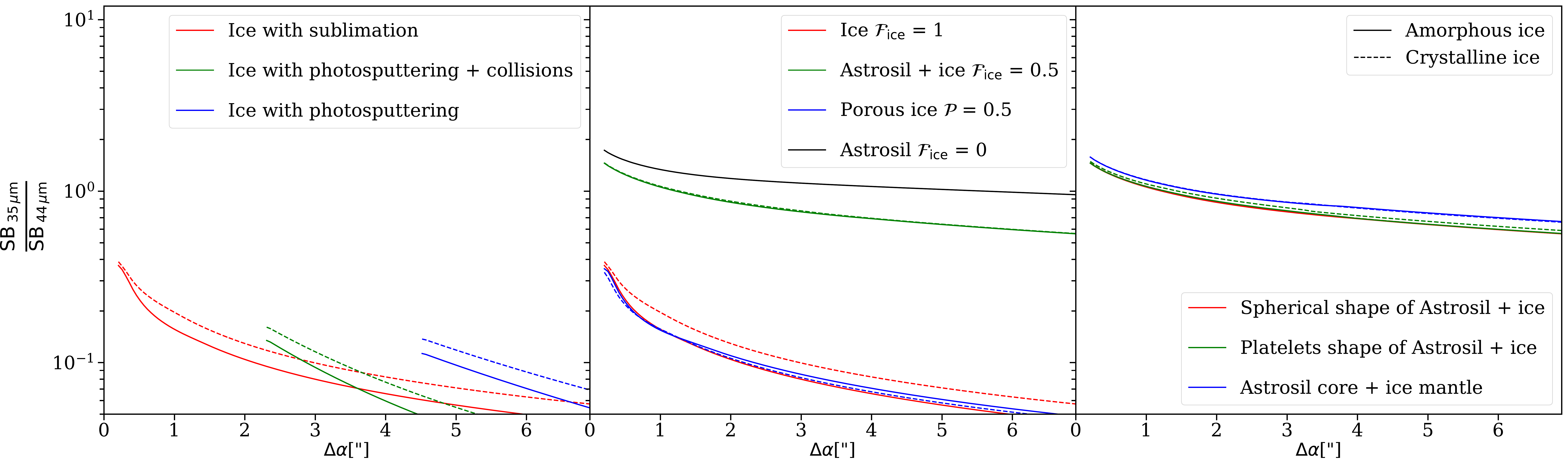}
\caption{Ratio between the surface brightness (SB) of debris disks assuming different mechanisms of ice destruction, chemical components, and shapes of icy-astrosilicate mixture at 35\,$\mu\rm m$ (i.e., outside of the 44\,$\mu\rm m$ ice feature) and 44\,$\mu\rm m$. The solid and dashed line indicate amorphous ice and crystalline ice, respectively.}
\label{FigVibStab}
\end{figure*}

{\hspace{-2mm}}\noindent \textbf{4.1.3. Different shape of icy dust aggregates} {\hspace{2mm}} In Fig. 10 we show the SED of debris disks that are composed of the icy dust of various shapes, that is, inclusion-matrix particles and core-mantle particles with spherical shape, inclusion-matrix particles with platelet shapes, and porous ice, assuming the same fractional ratio of ice (${\mathcal{F}}_{\rm ice}$~=~0.5). We find that the considered shapes affect the SED only weakly, except for the porous ice. However, the shape of icy dust aggregates matters for the strength of the ice features. Moreover, the ice features around 3~$\mu\rm{m}$, 44~$\mu\rm{m}$, and 62~$\mu\rm{m}$ as well as the astrosilicate feature 10~$\mu\rm{m}$ are weakly pronounced only in the case of core-mantle and inclusion-matrix particles. 
\subsection{Spatially resolved images}

\noindent We now discuss the influence of individual dust parameters and destruction of ice grains on scattered light to thermal reemission observations, that is, on wavelength-dependent spatially resolved images and their radial profiles from near-IR to submm wavelengths (at wavelengths $\lambda_{\rm obs}$~=~2.2\,$\mu\rm m$, 3.5\,$\mu\rm m$, 10\,$\mu\rm m$, 44\,$\mu\rm m$, and 1000\,$\mu\rm m$). We finally analyze and quantify the feasibility of constraining the spatial distribution of the smallest grains in the innermost warm disk regions by using the prominent ice and astronomical silicate features in the near-IR to mid-IR bands and the cold disk regions using corresponding features in the far-IR bands. \newline

{\hspace{-2mm}}\noindent \textbf{4.2.1. Pure ice} {\hspace{2mm}} Fig. 11 shows radial profiles of simulated observations of spatially resolved disks considering different destructive mechanisms of crystalline and amorphous ice, that is, sublimation, collisions, and UV photosputtering (see the panels in Fig. A. 1. for simulated observations of spatially resolved disks). When UV photosputtering is assumed, the surface brightness of debris disks is dominated by the large grains in the outermost cold disk regions. This is because energetic UV photons efficiently penetrate the disks out to very large distances, which critically decreases the abundance of the smaller particles by UV photosputtering. At 44\,$\mu\rm m$, that is, at the crystalline ice features, the flux density for crystalline ice slightly exceeds that of amorphous ice. This trend is drastically changed at submm wavelengths, which is due to the lower emissivity of crystalline ice (see Fig. 3). At submm wavelengths, we find that the surface brightness of the outer parts are even slightly brighter if UV photosputtering is considered instead of sublimation only (and/or collisions; see the right columns in Figs. 11 and 12). This is because UV photosputtering (and/or collisional effect) can no longer contribute efficiently to the destruction of larger grains (see Sect. 4. 1). In addition, collisional activity again clearly improves the situation for smaller and warmer ice survival in the inner region of debris disks. \newline

{\hspace{-2mm}}\noindent \textbf{4.2.2. Icy-astrosilicate dust aggregates} {\hspace{2mm}} Fig. 12 shows the radial profiles of simulated observations of spatially resolved disks considering different fractional ratios of crystalline and amorphous ice ${\mathcal{F}}_{\rm ice}$ in the icy-astrosilicate dust aggregates (see the panels in Fig. A. 2. for simulated observations of spatially resolved disks). In the inner part of the debris disk system, where only astrosilicate grains are present because of the ice sublimation (i.e., the porous astrosilicate grain), the surface brightness between debris disks with icy-astrosilicate dust mixtures and pure astrosilicate show a smaller difference. However, this difference increases toward the outer regions, where both astrosilicate and ice are present. This effect is more pronounced at 10\,$\mu\rm m$ observations, where we find an abrupt transition of the surface brightness around 40 - 60 au (see Fig. 14). This can be understood as a consequence of ice sublimation. In addition, we also find that the location of the change in surface brightness depends on the different physical state of icy-astrosilicate aggregates. This is due to the slight difference in the sublimation temperatures of both ice modifications in icy-astrosilicate aggregates (see Table 1). On the other hand, we find that the effect of the different fractional ratio of ice ${\mathcal{F}}_{\rm ice}$ on the surface brightness of debris disks at short wavelengths, that is, 2.2\,$\mu\rm m$ and 3.5\,$\mu\rm m$, and at submm wavelengths is weaker. This is because of the similar scattering and absorption cross section (C$_{\rm sca}$ and C$_{\rm abs}$) of ice and astrosilicate in the corresponding wavelength regime (see Fig. 3 and 4). The surface brightness of debris disks with pure crystalline ice at the submm wavelength is only significantly decreased due to their very different emissivity (see Fig. 1 and 3).\newline

{\hspace{-2mm}}\noindent \textbf{4.2.3. Different shape of icy dust aggregates} {\hspace{2mm}} Fig. 13 shows the radial profiles of simulated observations of spatially resolved disks considering different shapes of dust aggregates, that is, inclusion-matrix particles and core-mantle particles with spherical shape, and inclusion-matrix particles with platelet shapes with the same fractional ratio of ice, that is, ${\mathcal{F}}_{\rm ice}$~=~0.5 (see the panels in Fig. A. 3. for simulated observations of spatially resolved disks). The optical properties of ice-dust aggregates depend on size, shape, and physical states (amorphous vs. crystalline) of the dust grains. However, these differences are hardly noticeable in the flux density. Consequently, the spatially resolved disks images are hardly influenced by the different shape of the considered dust mixtures. \newline

\subsection{Prediction of ice reservoir location}

\noindent Based on the finding from Sect. 4. 2., we investigate the location of the ice survival line, focusing on grains of blowout size. Moreover, we study the radial position of the ice survival line as a function of various grain parameters such as physical state, the porosity of astrosilicate and ice, fractional ratio of ice ${\mathcal{F}}_{\rm ice}$, and shape of the dust aggregates. We note that the ice survival line also depends on the stellar luminosity. \newline 

{\hspace{-2mm}}\noindent \textbf{4.3.1. Prediction of ice reservoir location, the ice survival line} {\hspace{2mm}} In Fig. 14 we show the predicted location of the ice survival line for grains of blowout size. Depending on the chemical composition of dust aggregates with different fractional ratios of ice ${\mathcal{F}}_{\rm ice}$, the physical state of ice (amorphous or crystalline), its porosity, and the shape of the aggregates, we find that the ice survival line is located at $\text{about}$ 4.4 (for pure ice) - 26.3 (for icy-silicate aggregates) au from the host stars. This result is in good agreement with the previous study by \citet{Moerchen}, who found that the ice survival line of A-type star is located at $\text{about}$ 20 au. \newline
\indent The dependence on the physical state is due to the different emissivities (see also Fig. 3). Concerning the fractional ratio of ice ${\mathcal{F}}_{\rm ice}$, we find that the ice survival line is shifted toward the central star if the fractional ratio of ice ${\mathcal{F}}_{\rm ice}$ is decreased. Furthermore, thermal conductivity mainly depends on porosity. This means that larger porous dust grains are hotter than compact grains (\citealp{Krause}; \citealp{Kirchschlager13}; \citealp{Pawellek}; \citealp{Brunngraeber}). Consequently, for grains of a given size, porous grains are located farther out than compact grains. This moves the ice survival line to larger radii when the porosity is increased (see Fig. 14). In contrast to these findings, the shape of aggregates affects the location of ice survival line only weakly.\newline
 
{\hspace{-2mm}}\noindent \textbf{4.3.2. Spectral index $\alpha^{550\mu\rm{m}}_{2\rm{mm}}$} {\hspace{2mm}} The snow line causes a radial discontinuity in the spectral index profile (\citealp{Banzatti}). It imprints a strong signal on the dust thermal emission in the protoplanetary disks. Thus, we would expect to observe a similar phenomenon in debris disks. In Fig. 15 we show the spectral index, that is, $\alpha^{550\mu\rm m}_{2\rm mm}$, derived from spatially resolved simulations of the brightness profile of disks with ice with varying fractional ratios ${\mathcal{F}}_{\rm ice}$. We find a dependence of the discontinuity of the spectral index $\alpha^{550\mu\rm m}_{2\rm mm}$ on the fractional ratio of ice ${\mathcal{F}}_{\rm ice}$. It is located between~$\text{about}$~1.3 to 1.5 " (i.e., 25 to 30 au) from the star. A higher fractional ratio of ice ${\mathcal{F}}_{\rm ice}$ results in a slightly broader ring with a higher value of $\alpha^{550\mu\rm m}_{2\rm mm}$ (see Fig. 15). This is because a change in average grain size with a simultaneous change of composition from ice-astrosilicate aggregates to pure astrosilicate, resulting from the ice sublimation. Such particles show different absorption and emission behaviors than cold particles with ice. In addition, we find a similar phenomenon in the case of pure ice (red line in Fig. 15) at the inner part of debris disks. On the other hand, we find very different spectral indices in the case of the pure crystalline ice (the spectral slope is much steeper), resulting from the very low emissivity of pure crystalline ice (see Figs. 1, 3, and 5). 

\subsection{Evaluating the detectability of ice dust grains in future observations}
\noindent Finally, we use the spatially resolved images and radial profiles (Sect. 4.2) to predict the feasibility of detecting the spatially resolved characteristic structures with future observations such as the JWST/NIRCam and SPICA/SAFARI. \newline

{\hspace{-2mm}}\noindent \textbf{3\,$\mu\rm m$ H$_{\rm 2}$O ice band}{\hspace{2mm}} The NIRCam at the JWST, operating in the 0.6 to 5 $\mu\rm m$ wavelength range, consists of two modules (short-wavelength channel; 0.6 - 2.3 $\mu\rm m$ and long-wavelength channel; 2.4 - 5.0 $\mu\rm m$) that point to adjacent fields of view on the sky. The strong scattering feature of ice at around 2.8\,$\mu\rm m$ is located at this wavelength range (local minimum; see Fig. 1 and 2). In addition, the dependence on the particle phase is particularly high at $\text{about}$~3.5~$\mu\rm m$ (local maximum; see Figs. 1 and 2), which indicates possible candidate wavelengths to be compared.\newline
\indent Fig. 16 shows the ratio between surface brightness of debris disks with a different mechanism of ice destruction (left figure), chemical component (middle figure), and shape of icy dust mixture (right figure) inside and outside the 3\,$\mu\rm m$ ice feature. We find that the ratio is affected by the fractional ratio of ice ${\mathcal{F}}_{\rm ice}$. Ice-poor aggregates show a higher surface brightness ratio. The surface brightness is higher in the inner part of debris disks at 3.2\,$\mu\rm m$ (and in the outer part of debris disks at 2.8\,$\mu\rm m$) in the case of the icy-astrosilicate mixture and pure astrosilicate. An extreme increase in surface brightness ratio is therefore expected in the inner region (see the middle plot of Fig. 16). However, we find that the surface brightness ratio is almost constant and shows similar values in the case of pure ice for different ice destruction mechanisms (see the left plot of Fig. 16). This means that the ratio is no longer affected by the various depletion mechanisms. In addition, the porosity of ice (see the middle plot of Fig. 16) and the shape of the dust (see the right plot of Fig. 16) do not significantly affect the surface brightness ratio. Consequently, this comparison study allows constraining the existence of ice and even the fractional ratio of ice ${\mathcal{F}}_{\rm ice}$. \newline

{\hspace{-2mm}}\noindent \textbf{44\,$\mu\rm m$ H$_{\rm 2}$O ice band}{\hspace{2mm}} The SAFARI at SPICA will cover the far-IR window that extends from $\sim$ 34 $\mu\rm m$ to $\sim$ 230 $\mu\rm m$ with a field of view of 2'~$\times$~2' (\citealp{Roelfsema}). Thus, it will be possible to perform observations with medium spectral resolution over the solid-state ice features at $\sim$ 44 $\mu\rm m$ and 62 $\mu\rm m$ (see Fig. 1 and 2). These far-IR features will be useful for ice detection because the far-IR ice bands (broad features due to intermolecular lattice vibrations) are not confused with other solid-state features of less abundant species (unlike the mid-IR features, e.g., stretching, bending, or twisting of intramolecular bonds; \citealp{SPICA}). \newline 
\indent Fig. 17 shows the ratio between surface brightness of debris disks with a different mechanism of ice destruction (left figure), chemical component (middle figure), and shape of the icy dust mixture (right figure) inside and outside of 44\,$\mu\rm m$ ice feature. We find that the fractional ratio of ice determines the surface brightness ratio, for example, the existence of ice-poor aggregates causes a lower surface brightness ratio. In particular, the surface brightnesses in the two bands are nearly identical, that is, the ratio is close to 1 over the entire disk in the case of icy-astrosilicate aggregates and pure astrosilicate. However, because of the higher flux at the 44 \,$\mu\rm m$ ice feature, this effect is less pronounced in the case of pure ice. The ratio is below 1 and decreases significantly with increasing radial direction from the star. In addition, similar to the finding from Sect. 4.4.1, the different shapes of dust and the porosity of ice hardly affect the surface brightness ratio.


\section{Summary}

\noindent We investigated the feasibility of detecting water ice in typical debris disk systems assuming ice destruction mechanisms (sublimation of ice, dust production through planetesimal collisions, and photosputtering by UV bright central stars) and dust mixtures with various shapes consisting of amorphous ice, crystalline ice, astrosilicate, and vacuum inclusions. For this purpose, we first discussed the influence of these parameters on the resulting the SED (Section 4.1), spatially resolved images, and their radial profile (Section 4.2). Subsequently, we estimated and analyzed the location of the ice survival line as a function of these parameters (Section 4.3). Finally, we discussed the feasibility of detecting ice in debris disks in future observations (Section 4.4). Our key results are summarized below. \newline

\begin{enumerate}

        \item The sublimation of icy dust grains, collisions between planetesimals, and photosputtering due to UV sources clearly affect the observational appearance of debris disk systems. At near-IR to mid-IR wavelengths, the scattered radiation is significantly decreased by the destruction of small ice grains by UV photosputtering or collisions. At far-IR wavelengths, the thermal radiation from the dust is also significantly decreased because of the erosion of ice by UV photosputtering or collisions even far beyond the ice survival line. At submm wavelengths the effect of UV photosputtering/collision is weaker. However, the physical state of ice shows a strong effect on the spectral index of the SED. Furthermore, UV photosputtering and collisions determine the strength of the ice features. \newline
        
        \item The IR flux in the range of $\text{about}$ 10~$\mu\rm{m}$~to~80~$\mu\rm{m}$   increases with increasing porosity. In contrast, flux decreases with increasing porosity at shorter or longer wavelengths.\newline

        \item We found enhanced scattered-light polarization levels in the 3\,$\mu\rm{m}$ ice band for ice-rich aggregates, that is, a high fractional ratio of ice or highly porous ice. This means that the measurement of the wavelength-dependent polarization degree allows constraining the existence of ice or even the composition of icy dust grains. \newline
      
    \item The optical properties of dust grains depend on size, shape, and physical states (amorphous vs. crystalline) of the dust grains. However, these differences are hardly noticeable in the surface brightness scale of spatially resolved observations at K and L bands. At 10\,$\mu\rm{m}$, we find the abrupt transition of surface brightness as a consequence of ice sublimation, which depends on the fractional ratio of ice ${\mathcal{F}}_{\rm ice}$. At the submm wavelength, the surface brightness of debris disks with pure crystalline ice is only significantly decreased as a result of their very different emissivity. \newline      
      
    \item The radial position of the ice survival line depends on various grain parameters such as grain size, physical states, the porosity of ice, the chemical component with the different fractional ratio of ice, and the different shapes of aggregates. In the considered model, it covers a range from~$\text{about}$~4.4~to~26.3~au.\newline
      
    \item We discussed approaches to detect water ice grain with future observations with instruments operating in the near- to mid-IR (JWST/NIRCam) and far-IR (SPICA/SAFARI). \newline

\end{enumerate}


\begin{acknowledgements}
    We would like to thank the anonymous referee for helpful and constructive suggestions and comments that greatly contributed to improving the final version of the paper. This work was supported by the Research Unit FOR 2285 "Debris Disks in Planetary Systems" of the Deutsche Forschungsgemeinschaft (DFG). MK and SW acknowledge financial support under contracts WO~857/15-1~(DFG). AP and CJ acknowledge financial support under contracts JA~2107/3-1~(DFG). HM acknowledges financial support under contracts MU~1164/9-1~(DFG). We gratefully acknowledge the support by Volker Ossenkopf-Okada for making his \textbf{emc} tool for calculation of the optical properties of composite grains available. 
\end{acknowledgements}



\nocite{*}
\bibliographystyle{aa}
\bibliography{bibliography}


\clearpage
\appendix
\section{}

\noindent We present simulated observations of spatially resolved disks considering different ice dust parameters. The panels in Fig. A. 1. show simulated observations of spatially resolved disks considering different destructive mechanisms of crystalline and amorphous ice, that is, sublimation, collisions, and UV photosputtering (see Fig. 11 for the radial profile). The panels in Fig. A. 2. show simulated observations of spatially resolved disks considering different fractional ratios of crystalline and amorphous ice ${\mathcal{F}}_{\rm ice}$ in the icy-astrosilicate dust aggregates (see Fig. 12 for the radial profile). The panels in Fig. A. 3. show  radial profiles of simulated observations of spatially resolved disks considering different shapes of dust aggregates, that is, inclusion-matrix particles and core-mantle particles with spherical shapes, and inclusion-matrix particles with platelet shapes with the same fractional ratio as ice, that is, ${\mathcal{F}}_{\rm ice}$~=~0.5 (see Fig. 13 for the radial profile). 
\begin{figure*}
\centering
\includegraphics[width=18cm]{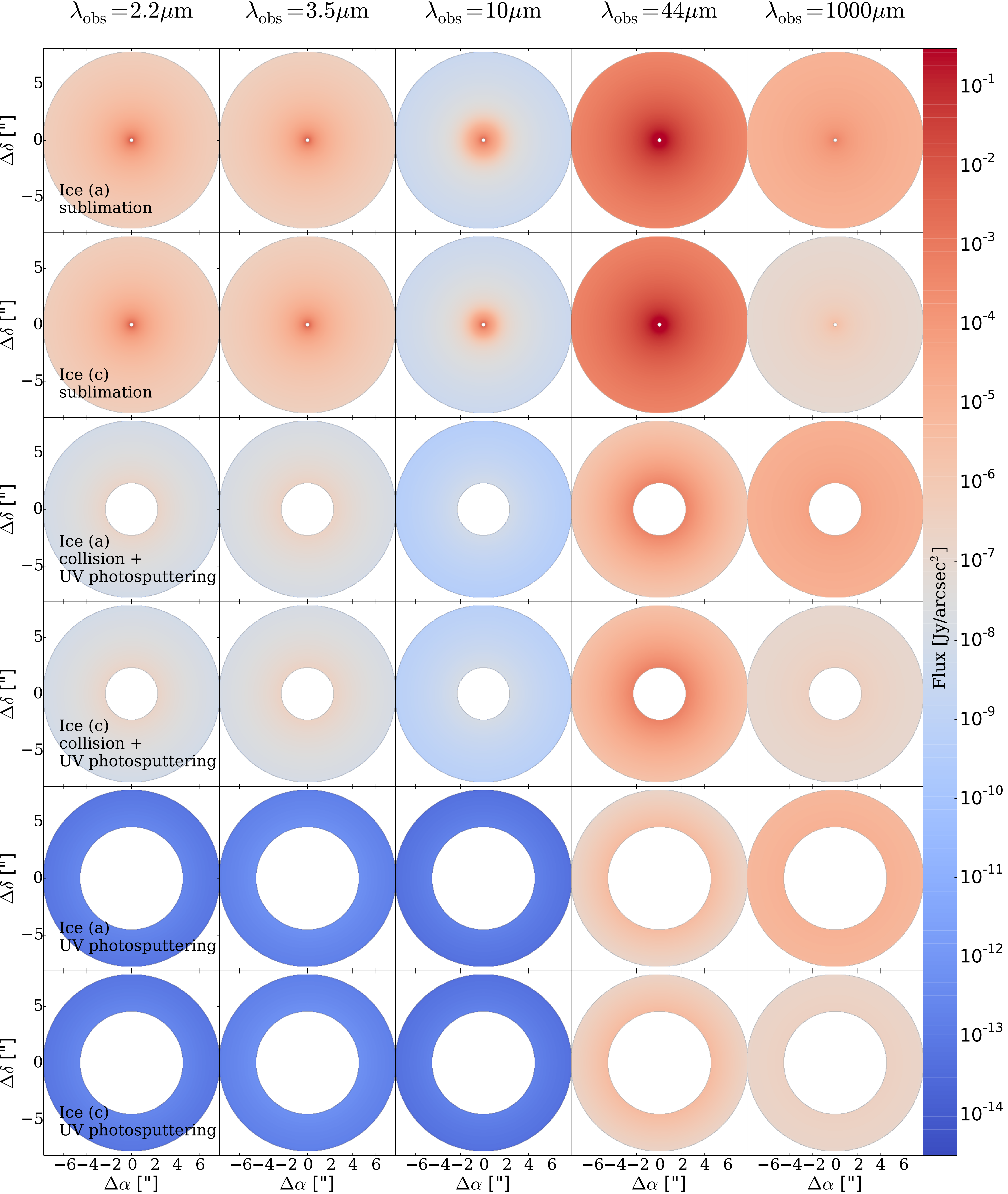}
\caption{Simulated surface brightness with debris disks composed of pure ice from near-IR to submm wavelengths at $\lambda_{\rm obs}$~=~2.2\,$\mu\rm m$, 3.5\,$\mu\rm m$, 10\,$\mu\rm m$, 44\,$\mu\rm m$, and 1000\,$\mu\rm m$. Different mechanisms of ice destruction are considered (indicated in each row). Ice (a) and ice (c) indicate amorphous and crystalline ice, respectively.}
\label{FigVibStab}
\end{figure*}
\begin{figure*}
\centering
\includegraphics[width=18cm]{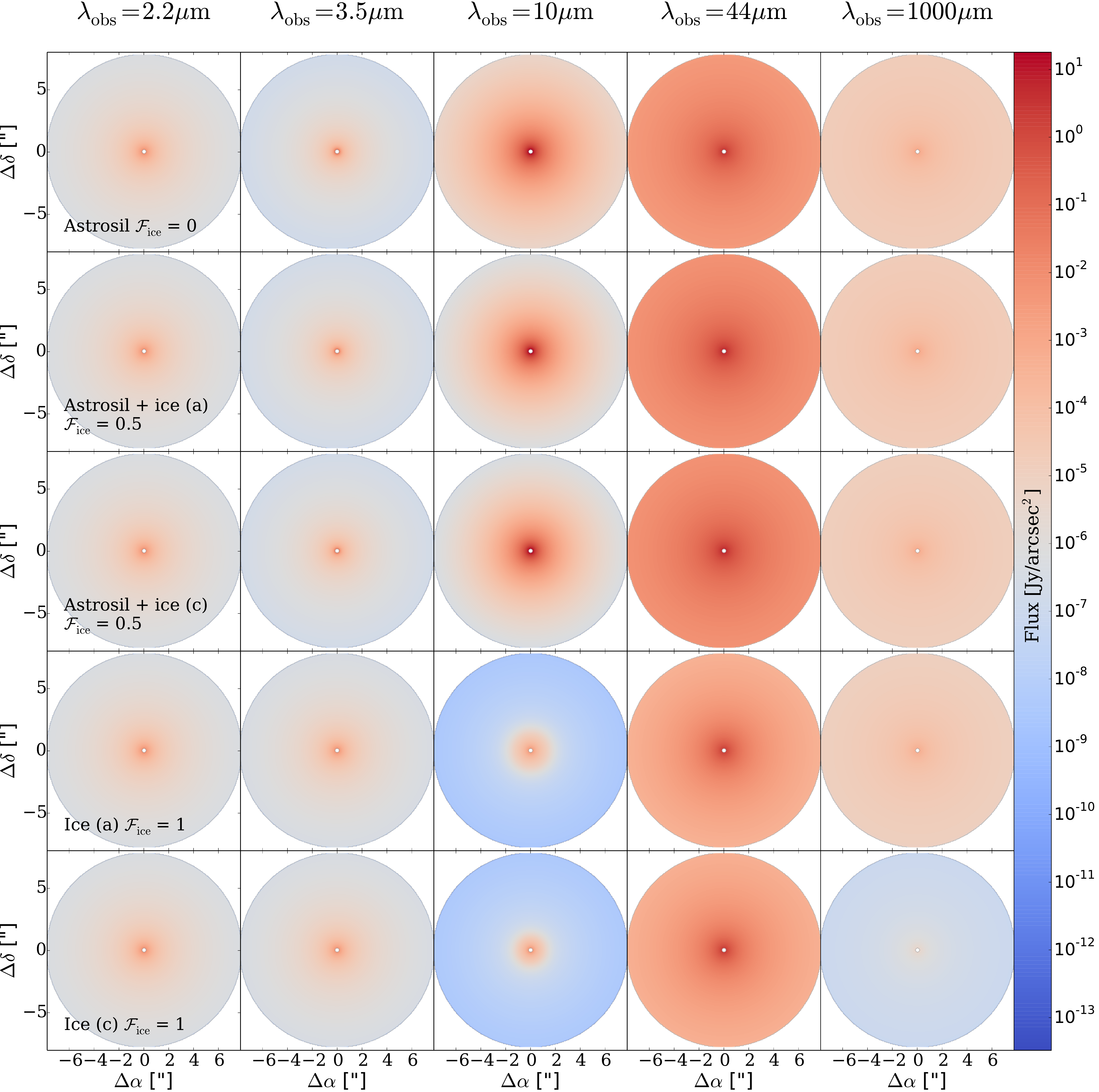}
\caption{Simulated surface brightness with debris disks composed of the icy dust mixture from near-IR to submm wavelengths at $\lambda_{\rm obs}$~=~2.2\,$\mu\rm m$, 3.5\,$\mu\rm m$, 10\,$\mu\rm m$, 44\,$\mu\rm m$, and 1000\,$\mu\rm m$. Different fractional ratios of ice ${\mathcal{F}}_{\rm ice}$ are considered (indicated in each row). Ice (a), ice (c), and astrosil indicate amorphous ice, crystalline ice, and astrosilicate, respectively.}
\label{FigVibStab}
\end{figure*}
\begin{figure*}
\centering
\includegraphics[width=18cm]{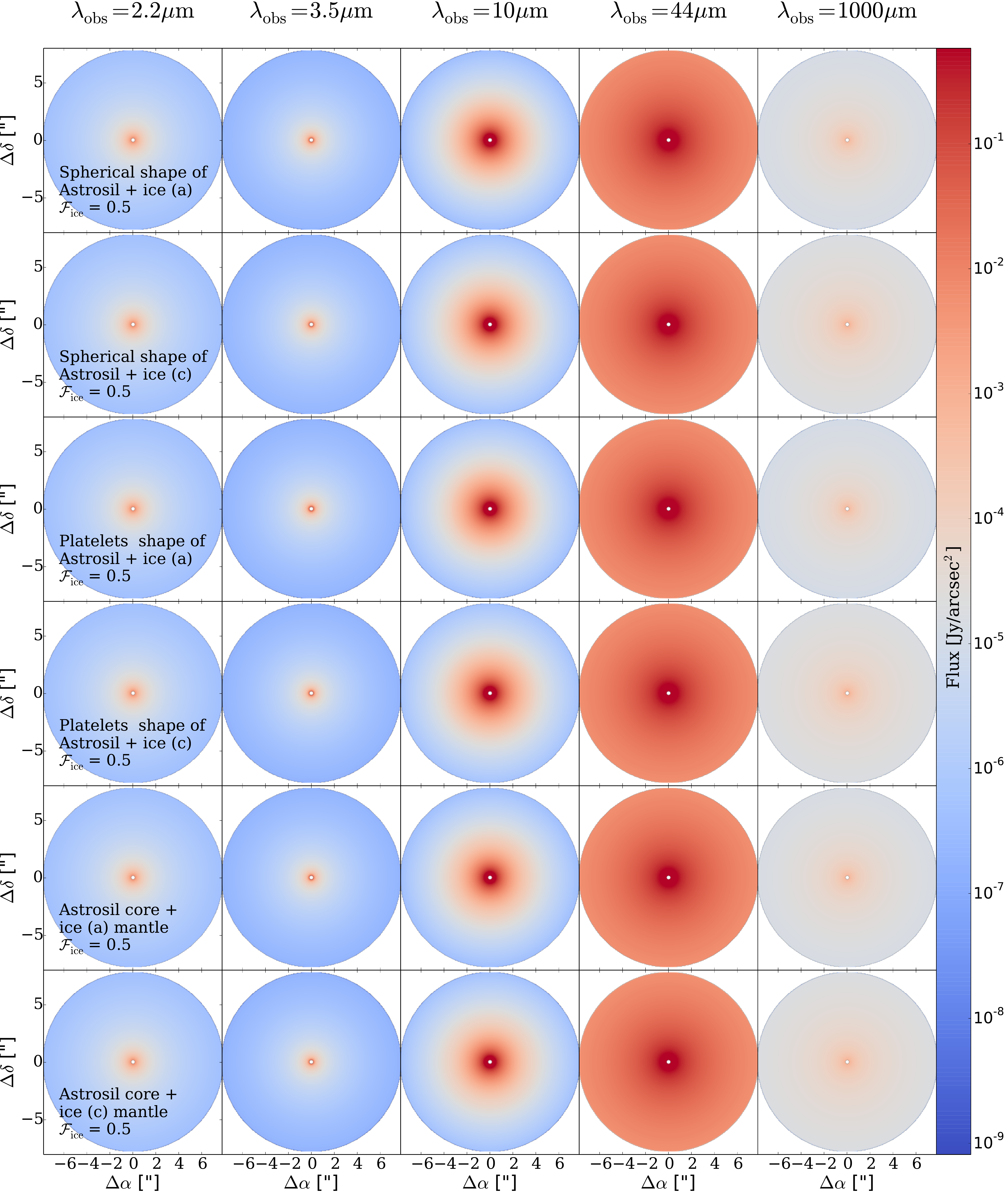}
\caption{Simulated surface brightness with debris disks composed of the icy dust mixture from near-IR to submm wavelengths at $\lambda_{\rm obs}$~=~2.2\,$\mu\rm m$, 3.5\,$\mu\rm m$, 10\,$\mu\rm m$, 44\,$\mu\rm m$, and 1000\,$\mu\rm m$. Different shapes of icy dust aggregates with the same ${\mathcal{F}}_{\rm ice}$~=~0.5 are considered (indicated in each row). Ice (a), ice (c), and astrosil indicate amorphous ice, crystalline ice, and astrosilicate, respectively.}
\label{FigVibStab}
\end{figure*}

\end{document}